\documentclass[aps,prl,groupedaddress,superscriptaddress,twocolumn,showpacs]{revtex4-1}

\usepackage{graphicx}
\usepackage{dcolumn}
\usepackage{bm}
\usepackage[T1]{fontenc}
\usepackage[french]{babel}
\usepackage{epstopdf}
\usepackage{amssymb}
\usepackage{color}
\usepackage{amsmath}
\usepackage{subfigure}
\usepackage{natbib}

\begin{document}

\preprint{APS/123-QED}

\title{Direct observation of long-range field-effect from gate-tuning of non-local conductivity}

\author{Lin Wang}
\author{Ignacio Guti\'{e}rrez-Lezama}
\affiliation{Department of Quantum Matter Physics, University of Geneva, 24 quai Ernest-Ansermet, CH-1211 Geneva, Switzerland}
\affiliation{Group of Applied Physics, University of Geneva, 24 quai Ernest-Ansermet, CH-1211 Geneva, Switzerland}

\author{C\'{e}line Barreteau}
\affiliation{Department of Quantum Matter Physics, University of Geneva, 24 quai Ernest-Ansermet, CH-1211 Geneva, Switzerland}

\author{Dong-Keun Ki}
\affiliation{Department of Quantum Matter Physics, University of Geneva, 24 quai Ernest-Ansermet, CH-1211 Geneva, Switzerland}
\affiliation{Group of Applied Physics, University of Geneva, 24 quai Ernest-Ansermet, CH-1211 Geneva, Switzerland}

\author{Enrico Giannini}
\affiliation{Department of Quantum Matter Physics, University of Geneva, 24 quai Ernest-Ansermet, CH-1211 Geneva, Switzerland}

\author{Alberto F. Morpurgo}
\affiliation{Department of Quantum Matter Physics, University of Geneva, 24 quai Ernest-Ansermet, CH-1211 Geneva, Switzerland}
\affiliation{Group of Applied Physics, University of Geneva, 24 quai Ernest-Ansermet, CH-1211 Geneva, Switzerland}


\begin{abstract}
We report the direct observation of a long-range field-effect in WTe$_2$ devices, leading to large gate-induced changes of transport through crystals much thicker than the electrostatic screening length. The phenomenon --which manifests itself very differently from the conventional field-effect-- originates from the non-local nature of transport in the devices that are thinner than the carrier mean free path. We reproduce theoretically the gate dependence of the measured classical and quantum magneto-transport, and show that the phenomenon is caused by the gate-tuning of the bulk carrier mobility by changing the scattering at the surface. Our results demonstrate experimentally the possibility to gate tune the electronic properties deep in the interior of conducting materials, avoiding limitations imposed by electrostatic screening.
\end{abstract}

\maketitle
Conventional field-effect transistors (FETs) exploit electrostatic gating to tune the electronic properties of materials by means of charge accumulation~\cite{Ahn2006,Ohno2000,Novoselov2004,Martel1998,Caviglia2008}. Gate-induced charge accumulation occurs close to the material surface, on a depth limited by the so-called screening length, which is typically very short, $\sim$1-2 nm. Electrostatic screening, therefore, seems to preclude the possibility to use FET devices to control the electronic properties in the interior of materials, i.e., their bulk response. Although this is indeed the case in conventional field-effect devices, here we report the observation of a much longer-range field-effect, affecting electronic transport through a material over a depth orders of magnitude longer than the electrostatic screening length. The phenomenon, which occurs because the electrical conductivity is governed by non-local processes, manifests itself in large gate-induced changes in the transport properties of conductors as long as their thickness is smaller than or comparable to the carrier mean free path.

We observe such a long-range field-effect in crystals of WTe$_2$, a material possessing remarkable electronic properties~\cite{Ali2014,Pletikosic2014,Jiang2015,Zhu2015,Rhodes2015,Zhao2015,Cai2015,Pan2015,Kang2015,Thoutam2015,Wu2015,Wang2015b,Dai2015,Homes2015,Kong2015,Das2016,Wang2015,Alekseev2015,Soluyanov2015,Qian2014}. Transport experiments have shown that bulk WTe$_2$ is a nearly perfectly compensated semi-metal exhibiting record-high magnetoresistance (MR) because of the high electron and hole mobility~\cite{Ali2014,Zhu2015,Wang2015}. They have also shown that whenever the crystal thickness is reduced below the mean-free path (few hundreds nanometers or even longer), the carrier mobility is suppressed by scattering at the surface~\cite{Wang2015}. As established long ago, this implies that transport at the microscopic scale is governed by non-local processes, i.e. the relation between current density and electric field is non-local~\cite{Sondheimer1952,Schrieffer1955,Price1960,Ibach2006,Fu2016,Fu2016a}~\footnote{The non locality that we are referring to is described by a relation between current density $\vec{J}$ and electrical field $\vec{E}$, of the type $\vec{J}(x)=\int dy\sigma(x,y)\vec{E}(y)$}. It is well-known that in this non-local regime different physical phenomena exhibit an unusual behavior, as illustrated by so-called anomalous skin effect~\cite{Reuter1948,Chambers1950,Chambers1952}, i.e. the possibility for electromagnetic waves to penetrate into a conductor over a distance much larger than that predicted by the conventional theory. Although in the past it had been realized that gate-tuning of surface scattering could result in gate dependence of transport properties in systems --such as metals-- in which no field-effect should be expected~\cite{Berman1971,Berman1975}, no direct experimental demonstration of this phenomenon and of its long-range nature has been provided~\footnote{In Refs.~\cite{Berman1971,Berman1975} the observation of a (gate-induced) 10$^{-5}$ relative change in the conductivity of metallic thin films was claimed to represent a manifestation of these phenomena. However, no actual evidence for a long-range field-effect was presented, as the claim was entirely based on data interpretation (using the phenomenological Fuchs-Sondheimer model) and not on direct measurements. It is by now known that the assumptions made in Refs.~\cite{Berman1971,Berman1975} are wrong and so are the conclusions (thorough discussions of this issue can be found in numerous papers, including A. F. Mayadas \textit{et al.}, Appl. Phys. Lett. \textbf{14}, 345 (1969); A. F. Mayadas and M. Shatzkes, Phys. Rev. B \textbf{11}, 1382 (1970); R. C. Munoz \textit{et al.}, Phys. Rev. B \textbf{62}, 7 (2000))}.

Our devices consist of  WTe$_2$ crystals with thickness ranging from 10 to 50 nm exfoliated onto a highly doped Silicon substrate covered with a 285 nm SiO$_2$ insulating layer~\cite{Wang2015}. The doped Silicon substrate can be used as a gate electrode, even though for most devices --and certainly in the 50 nm thick crystals-- no significant gate-induced modulation of transport is \textit{a priori} expected. Indeed, applying a large gate voltage, $V_g=80$ V accumulates a charge of 8$\times$10$^{12}$ carriers/cm$^2$ at the surface, corresponding approximately to only 3-4 \% of the total amount of charge carriers present in a 50 nm thick crystal. The resulting modulation in conductivity is expected to be even much smaller, as in a semi-metal the gate voltage increases the surface density of one type of charge carriers and decreases that of the other, so that the effect on the conductivity largely compensates.

\begin{figure}[t]
\includegraphics[width=8.5cm]{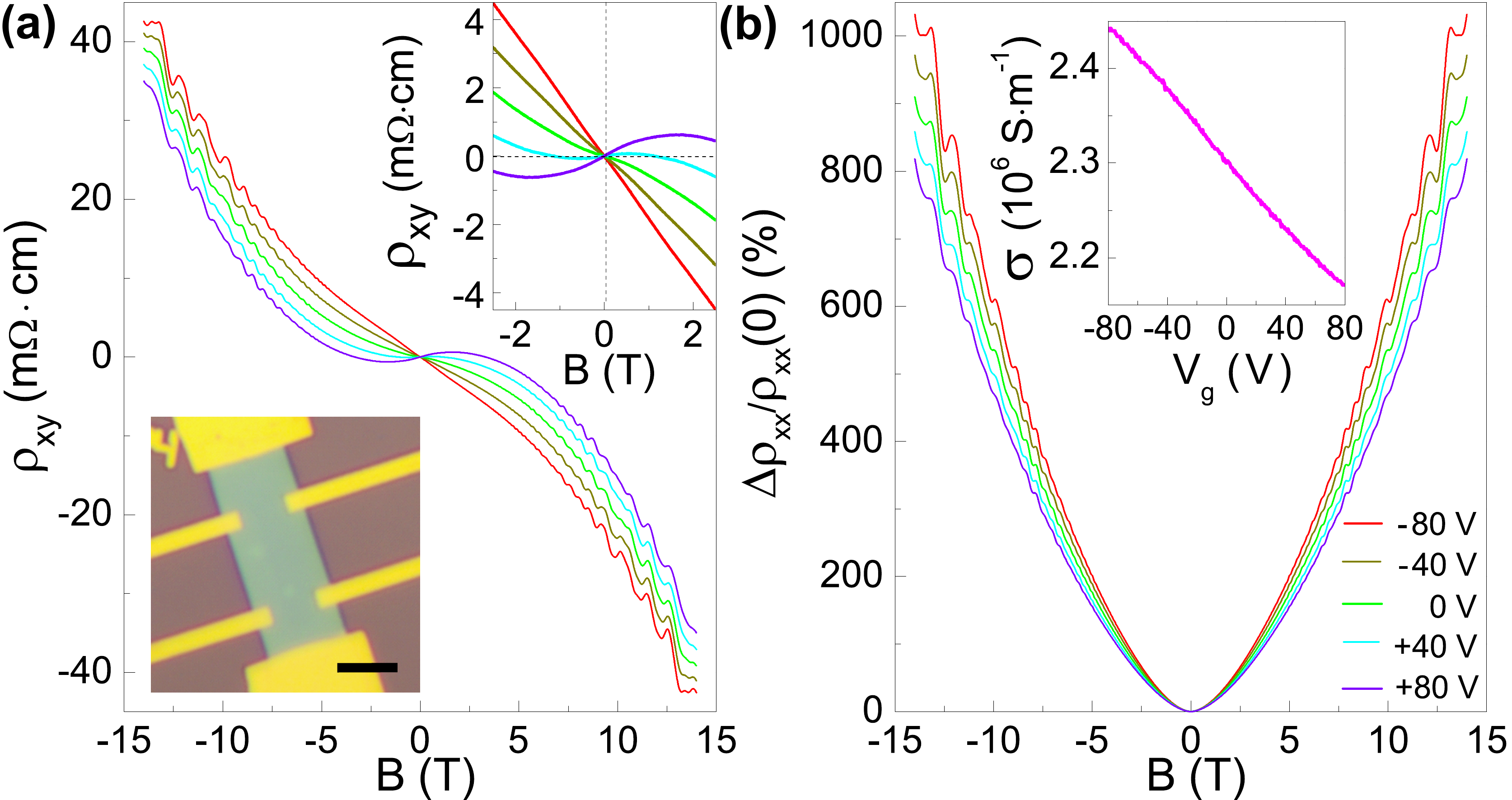}
\caption{\label{fig1} (Color online) Gate-induced modulation of magneto-transport in a 48 nm thick WTe$_2$ crystal (Sample A). (a) Transverse resistivity $\rho_{xy}$ showing a pronounced gate voltage $V_g$ dependence (the curves of different color correspond to different values of $V_g$, as shown by the legend in panel (b)). Note the change of sign occurring for $|B|<3$ T, enlarged in the upper inset. The bottom inset shows an optical microscope image of the device (the bar is 5 $\mu$m). (b) $V_g$ dependence of the longitudinal magnetoresistance (MR) of the same device. The inset shows the conductivity $\sigma(V_g)$ measured at $B=0$ T. All data were taken at $T=250$ mK.}
\end{figure}

At odds with these expectations, Fig. 1 shows a pronounced effect of an applied $V_g$ already on a 48 nm thick WTe$_2$ crystal (hereafter referred to as Sample A). Fig. 1(a) shows that the modulation in the Hall resistivity $\rho_{xy}$ is so large that the sign of $\rho_{xy}$ is inverted for magnetic field $B$ up to 2-3 T. Even more surprisingly, the evolution of the low-$B$ slope of $\rho_{xy}$ is not consistent with the sign of the charges accumulated by the gate. Specifically, at a positive $V_g>0$ V, electrons are accumulated, which should drive the $\rho_{xy}$ towards a negative slope. The inset of Fig. 1(a), however, shows the opposite behavior: the low-$B$ slope of $\rho_{xy}$ is negative at $V_g < 0$ V and becomes positive at $V_g = +80$ V. Furthermore, the conductivity $\sigma$ measured at $B=0$ T exhibits a $V_g$ dependence opposite to that naively expected. Since we know from the analysis of magneto-transport at $V_g=0$ V that in this device $\mu_e > \mu_h$ (see Ref.~\cite{Wang2015}), a positive $V_g$ --which increases the electron density $n$ and decreases the density of holes $p$-- should slightly increase the total conductivity $\sigma=ne\mu_e + pe\mu_h$ (where $\mu_e$ and $\mu_h$ are electron and hole mobility). However, the inset of Fig. 1(b) shows that $\sigma$ decreases upon driving $V_g$ more positive. Therefore, electrostatic gating of a rather thick WTe$_2$ crystal results in sizable changes of the transport properties that are entirely inconsistent with the effect expected due to the accumulated surface charges, i.e. with the behavior of conventional field-effect.

A hint to explain the observed gate-dependent behavior comes from the longitudinal MR measurements. Fig. 1(b) shows that the MR, while being modulated by $V_g$, keeps exhibiting a quadratic dependence on $B$, consistent with $\frac{\Delta\rho_{xx}}{\rho_{xx}}=\frac{\rho_{xx}(B)-\rho_{xx}(0)}{\rho_{xx}(0)}=\mu_e\mu_hB^2$~\cite{Wang2015}. This relation indicates that the MR depends only on $\mu_e$ and $\mu_h$, suggesting that the observed $V_g$ dependence of transport originates from a modulation of the mobility of the bulk carriers. This is possible because for all devices investigated here the electron and hole mean free paths ($L_e \sim L_h=\mu_{e,h}\hbar k_F/e$) are larger than the WTe$_2$ crystal thickness~\footnote{For Sample A, for instance, $\mu_e=2.900$ cm$^2$V$^{-1}$s$^{-1}$ and $\mu_h=2.900$ cm$^2$V$^{-1}$s$^{-1}$, so that the electron and hole mean-free paths $L_h\sim L_e\sim0.3$ $\mu$m $\gg t$; the Fermi vector $k_F$ is estimated from Refs.~\cite{Pletikosic2014,Jiang2015,Rhodes2015,Zhu2015}}, so that the carrier mobility in the bulk is non-locally determined by the scattering at the surface~\cite{Wang2015}.

To understand physically how gating can affect the mobility of bulk carriers, it is sufficient to look at the gate-induced bending of the valence and conduction band near the material surface (see Fig. 2). At $V_g=0$ V (Fig. 2(b)), the Fermi energy $E_F$ is located inside the overlapping conduction and valence band uniformly throughout the entire thickness of the crystal, all the way up to the surface next to the gate dielectric. Electrons and holes move freely in the bulk and can reach the surface, where --as we know from  past work~\cite{Wang2015}-- they undergo scattering processes that determine their mobility $\mu_e$ and $\mu_h$. Although the precise mechanism is yet unknown, all observations~\cite{Wang2015} indicate that the surface scattering is short-ranged and mainly affects electrons reaching the outermost layer. A negative $V_g < 0$ V (Fig. 2(c)) increases the electrostatic energy of electrons resulting --for sufficiently large $V_g$ values-- in their depletion next to the surface. Under these conditions, electrons do not have enough kinetic energy to reach the surface, and suffer therefore less scattering processes. As a result their mobility increases. The same logic applies to holes for a sufficiently large positive $V_g > 0$ V (Fig. 2(d)). We therefore expect that $\mu_e$ and $\mu_h$ should depend on $V_g$ and exhibit opposite trends as the gate voltage is varied.

\begin{figure}[t]
\includegraphics[width=8.5cm]{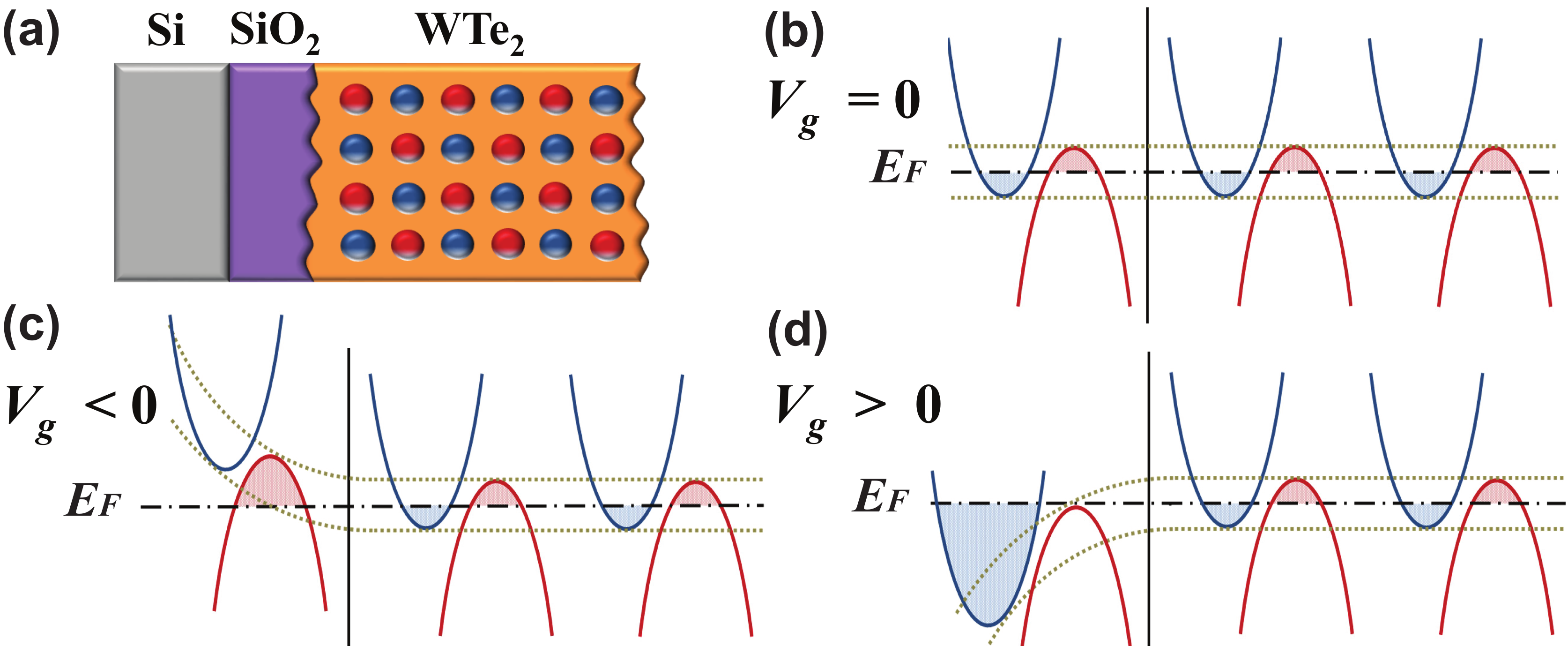}
\caption{\label{fig2} (Color online)  (a) Schematic illustration of the device structure (the blue and red balls in the WTe$_2$ layer represent electrons and holes). (b-d) Band bending for different values of $V_g$. For $V_g=0$ V (b) the system is uniform and electron and holes can reach the surface. For large negative $V_g$ (c), electrons cannot reach the surface --where scattering processes predominantly occurs-- and their mobility increases. (d) The same holds true for holes at large positive $V_g$.}
\end{figure}

To confirm the validity of this physical scenario we perform a complete quantitative analysis of the measured gate-dependent classical magneto-transport in terms of an electron/hole two-band  model. Changing  $V_g$ has a non-local and a local effect: it varies the mobility $\mu_{e,h}$ of electrons and holes in the bulk (non-local effect) without changing their density $n$ and $p$, and it changes the density of charge carriers at the surface (within the electrostatic screening length, $\approx 1$ nm, much smaller than the crystal thickness; local effect), which can also cause changes in the magneto-transport. In terms of the longitudinal and transverse square conductance $G_{\square,xx}$ and $G_{\square,xy}$ we then have :
\begin{align}
G_{\square,xx} &= \sigma_{xx,bulk}\cdot t + \sigma_{xx,inter}, \\
G_{\square,xy} &= \sigma_{xy,bulk}\cdot t + \sigma_{xy,inter},
\end{align}
where $\sigma_{bulk}$ ($\sigma_{inter}$) is the 3D (2D) bulk (interface) conductivity. As discussed theoretically long ago~\cite{Sondheimer1952,Schrieffer1955}, in the non-local transport regime occurring because of the presence of surface scattering, the carrier mobility $\mu_{e,h}$ and the (bulk) conductivity are defined as an average over crystal thickness $t$ (e.g., $\mu_{e,h}=\frac{1}{t}\int_0^t \mu_{e,h}(z) dz$). It is the introduction of these effective, averaged quantities that accounts for the non locality of the relation between current density and electric field, which is at the core of the phenomenon observed here (see Ref.~\cite{Sondheimer1952,Schrieffer1955,Price1960,Ibach2006}).

We discuss in detail the behavior of thick crystals in which the surface contribution, $\sigma_{xx,inter}$ and $\sigma_{xy,inter}$, can be entirely neglected with respect to the bulk one. This allows us to minimize the number of unknown parameters in the data analysis (the behavior of thinner crystals can also be reproduced in detail, as discussed in detail in the Supplemental Material~\cite{support}). By using the two-band model expressions for the bulk electron and hole classical conductivities Eqs. (1) and (2) reduce to~\cite{Murzin1946,Sondheimer1947}:
\begin{align}
G_{\square,xx} &= \sigma_{xx,bulk}\cdot t = \left(\frac{pe\mu_h}{1+\mu_h^2B^2}+\frac{ne\mu_e}{1+\mu_e^2B^2}\right)t, \\
G_{\square,xy} &= \sigma_{xy,bulk}\cdot t = \left(\frac{pe\mu_h^2B}{1+\mu_h^2B^2}+\frac{ne\mu_e^2B}{1+\mu_e^2B^2}\right)t.
\end{align}

\begin{figure}[t]
\includegraphics[width=8.5cm]{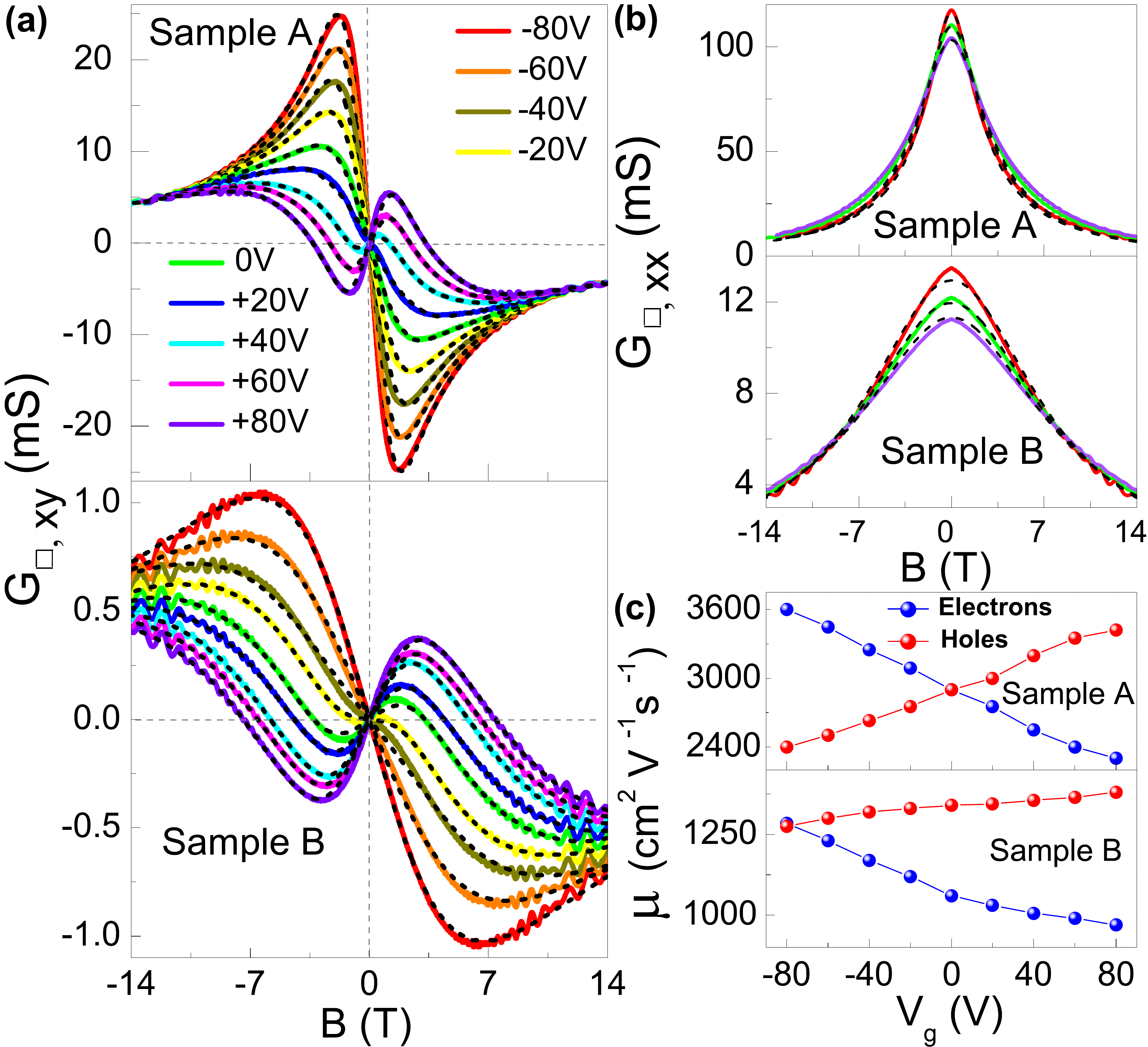}
\caption{\label{fig3} (Color online) Quantitative analysis of the gate-dependent magneto-transport through WTe$_2$ crystals of different thickness. (a-b) Transverse and longitudinal square conductance, $G_{\square,xy}(B)$ and $G_{\square,xx}(B)$, of a 48 nm (Sample A) and an 11 nm thick (Sample B) crystal measured for different $V_g$ at $T=250$ mK. In all panels, curves of the same color correspond to the same value of $V_g$, as indicated by the legend in (a). The black dashed lines represent theoretical fits with Eqs. (1-4), which reproduce the data quantitatively in all detail. (c) $V_g$-dependence of the electron (blue symbols) and hole (red symbols) mobility extracted from the analysis.}
\end{figure}

We first extract the (bulk) electron and hole density by fitting $G_{\square,xx}(B)$ and $G_{\square,xy}(B)$ at $V_g = 0$ V (see Ref.~\cite{Wang2015} for details). Then --as we change $V_g$-- we keep $n$ and $p$ fixed to the value determined at $V_g=0$ V and vary only  $\mu_e$ and $\mu_h$, to reproduce $G_{\square,xx}(B)$ and $G_{\square,xy}(B)$. As shown in Figs. 3(a-b), the agreement between Eqs. (3-4) and the data is excellent throughout the $V_g$ and $B$ ranges investigated. In particular, Eq. (4) very successfully reproduces the non-trivial evolution of $G_{\square,xy}$ including its sign changes and the inversion of the slopes at low B. The values of $\mu_e(V_g)$ and $\mu_h(V_g)$ extracted from the fitting are plotted in Fig. 3(c), and exhibit the trends expected from the proposed physical scenario. The electron mobility increases as $V_g$ becomes more negative, i.e., when electrons are pushed away from the interface, whereas the hole mobility exhibits the opposite behavior. The total change in either the electron or the hole mobility is less than a factor of two, as it should: even if scattering at one surface is fully suppressed, the non-gated surface continues to limit the mobility. A similar quantitative analysis on 6 devices (out of more than 20 devices measured which exhibited the same trends) resulted in all cases in excellent agreement with Eqs. (1-4) and identical trends for the $V_g$ dependence of $\mu_{e,h}$ (see Supplemental Material~\cite{support} for details). We therefore conclude that the observed unusual gate-induced variations of transport are caused by changes in the mobility of the bulk electrons and holes, and occur because of the non-local transport regime in which the devices operate~\footnote{This conclusion is confirmed by the observation, discussed in the Supplemental Material~\cite{support}, that when the exposure to ambient of the WTe$_2$ crystals is only very limited --so that only minor surface degradation occurs-- no gate-voltage dependence of transport is observed experimentally. In general the magnitude of the field-effect that we observe depends on the amount of surface degradation which is determined by details of the fabrication process.}.

Such a field-effect mechanism had not been directly observed previously [38]. In semi-metallic graphite or bismuth, for instance, a gate modulation of transport is routinely found in crystals with thickness up to a few tens of nanometers~\cite{Zhang2005,Zhang2005a,Butenko1997,Butenko1999}. In that case, however, the modulation of transport is due to the contribution to the conductivity given by the carriers accumulated near the surface, and for 50 nm or thicker crystals the effect is virtually negligible. In other kinds of transistors, a modulation of transport due to a gate-induced change in carrier mobility originating from the effect of surface roughness has been well documented~\cite{Ando1982,Ahn2006}. In those transistors, however, carriers form a 2D conducting layer confined near the material surface, and the gate voltage does not influence the electronic properties in the interior of the material. The unique aspect of the field-effect observed in our study of WTe$_2$ devices is that the gate voltage has an influence on the electronic properties over the entire material even for rather thick crystals.

\begin{figure}[t]
\includegraphics[width=8.5cm]{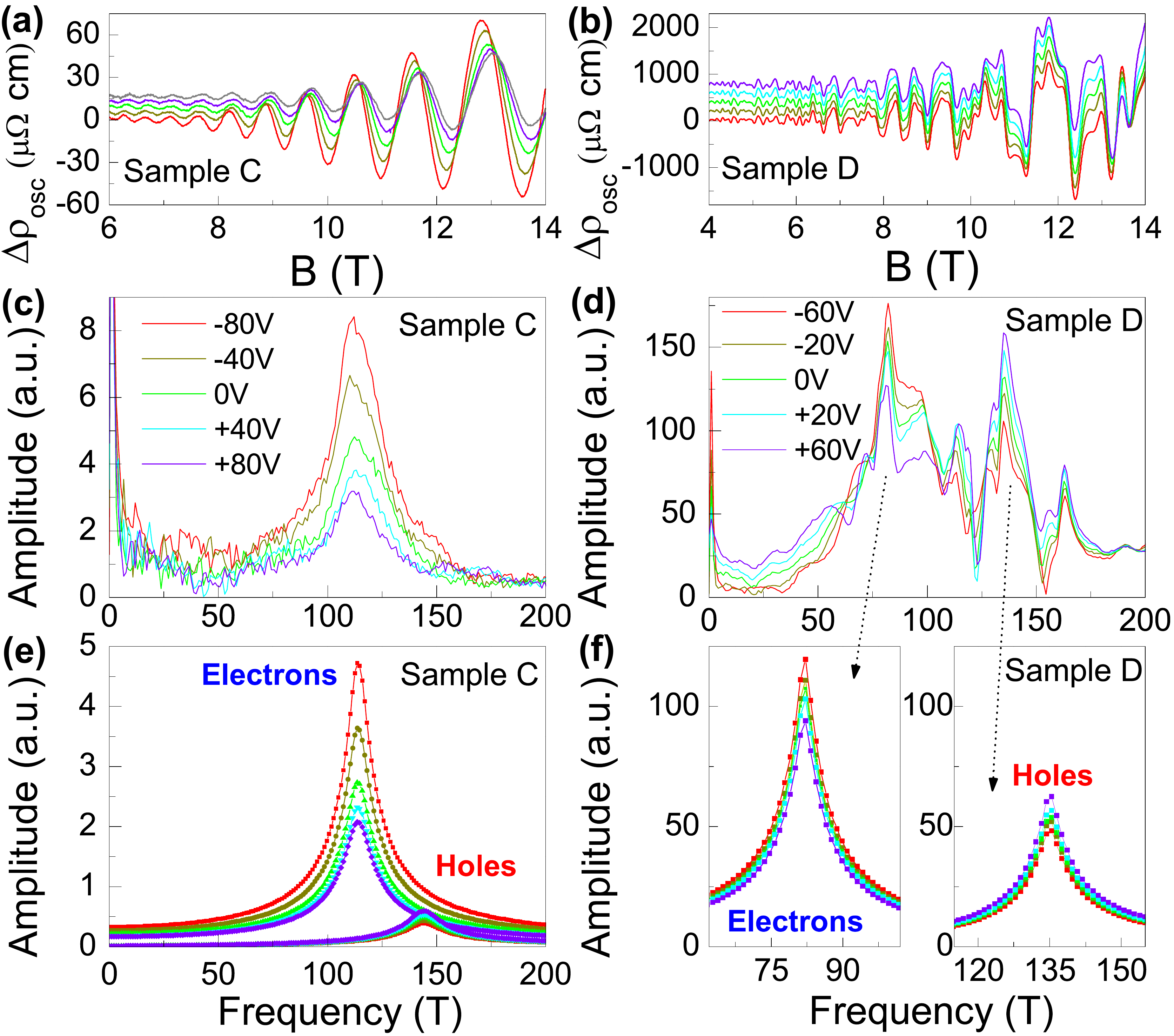}
\caption{\label{fig4} (Color online) Gate-dependent Shubnikov-de Haas (SdH) oscillations. (a-b) Oscillatory component of the resistivity $\Delta \rho_{osc}$ --measured at $T=250$ mK-- for different values of $V_g$ for two different devices, based on (a) a 11 nm (Sample C) and (b) a 37 nm thick (Sample D) WTe$_2$ crystals (the data have been offset for clarity). (c-d) Fourier spectrum of the SdH oscillations shown in panels (a-b). (e-f) Fourier spectrum of the SdH oscillation $\Delta \rho_{osc}^{calc}$ calculated using Eq. (5) with the values of $\mu_e(V_g)$ and $\mu_h(V_g)$ extracted from the analysis of classical magneto-transport. Curves with the same color in panels (a,c,e) and (b,d,f) are taken at the same $V_g$ indicated by the legends in panels (c) and (d), respectively.}
\end{figure}

To illustrate why the ability to gate-tune the bulk properties of WTe$_2$ is particularly interesting, we discuss the effect of $V_g$ on the (quantum) Shubnikov-de Haas (SdH) oscillations originating from the formation of Landau levels in the bulk. The conventional theoretical expression describing the oscillatory component of $\rho_{xx}$, $\Delta\rho_{osc}$ reads~\cite{Richards1973,Niederer1974,Eisele1976,Fang1977,Bangura2008}:
\begin{equation}
\frac{\Delta\rho_{osc}}{\rho_{xx}} \propto \sqrt{\frac{\hbar eB}{m^*E_F}}\frac{X}{\sinh X}\exp\left(\frac{-\pi}{\mu B}\right)\sin\left(\frac{2\pi f}{B}\right),
\label{SdH}
\end{equation}
where $X=\frac{2\pi^2k_BTm^*}{\hbar eB}$, $k_B$ is the Boltzmann constant, $T$ the temperature, and $f$ the oscillation frequency. From the analysis of classical transport, all parameters are known and we can use Eq. (5) to calculate the evolution of the SdH oscillations with $V_g$~\footnote{For the electron/hole effective masses we use $m_e^*=0.5m_0$, $m_h^*=1.0m_0$, in the range of values reported in the literature~\cite{Pletikosic2014,Jiang2015,Rhodes2015,Zhu2015}; $m_e^*= 0.33\sim0.51 m_0$ and $m_h^*= 0.42\sim1.1 m_0$, with $m_0$ the free electron mass; for $f$, we insert the observed value}. We then compare the Fourier spectrum obtained from the measured oscillations $\Delta \rho_{osc}$ to that from the calculated $\Delta \rho_{osc}^{calc}$ by using Eq. (5) in the same $B$-range of the measurements. Results for two different devices are shown Figs. 4(a-d): Fig. 4(a) and (c) illustrate the behavior of a device realized on an $\sim$11 nm thick WTe$_2$ crystal with relatively low mobility, whereas Figs. 4(b) and (d) are from a high-mobility device ($\mu_{e,h}\sim5.000-6.000$ cm$^2$V$^{-1}$s$^{-1}$), whose behavior represents that of WTe$_2$ crystals that are 35-50 nm thick. The corresponding theoretical results are shown in Figs. 4(e-f).

Starting with the thin device, Fig. 4(c) at $V_g = 0$ V shows a single broad peak at $f \sim 114$ T, with a faint shoulder at $f \sim 144$ T (see also the data from Sample E shown in Supplemental Material~\cite{support}). Upon changing $V_g$, the positions of the peak and shoulder do not change, since the bulk density is unaffected by $V_g$. The amplitude of the peak, on the contrary, changes considerably, consistently with the change in carrier mobility. Since at more positive $V_g$ $\mu_e$ decreases and $\mu_h$ increases, the strong suppression with increasing $V_g$ indicates that the peak originates from SdH oscillations of electrons. For comparison, Fig. 4(e) shows the spectrum obtained from Eq. (5) using the values of $\mu_e(V_g)$ and $\mu_h(V_g)$ extracted from the analysis of classical transport on the same device. The trend of the theoretically calculated and experimentally measured curves match satisfactorily: electron SdH oscillations have a stronger $V_g$ dependence, while hole SdH oscillations (responsible for the shoulder at $f \sim 144$ T; see Supplemental Material~\cite{support}) exhibit a much smaller $V_g$-dependence due to their lower mobility and larger effective mass. Most typically, especially in the thin devices, $\mu_e > \mu_h$ and SdH oscillations exhibit predominantly the $V_g$ dependence expected for electrons.

In the highest mobility devices, however, both electron and hole SdH oscillations are clearly visible in the experiments (Fig. 4(d)). The evolution of the spectrum with increasing $V_g$ from -60 V to 60 V is opposite in different frequency ranges. For 75 T $<f<$ 105 T, the spectrum amplitude decreases upon increasing $V_g$, consistently with SdH oscillations caused by electrons, whereas for 115 T $<f<$ 165 T the opposite trend is clearly visible, as expected for holes. The fine structure present in the spectrum (usually seen in devices having this mobility and thickness, and possibly originating from size quantization generating multiple electron and hole sub-bands) prevents a quantitative comparison. Nevertheless, with $\mu_e(V_g)$ and $\mu_h(V_g)$ extracted from the classical magneto-transport analysis, Eq. (5) predicts that the relative variations in the electron and hole contributions induced by $V_g$ are comparable in magnitude (Fig. 4(f)), as found in the experiments. We conclude that --for sufficiently high mobility devices-- the gate dependence allows the identification of the carriers responsible for the SdH oscillations observed at a certain frequency.

In conclusion, we have observed and explained a long-range field-effect of magneto-transport in WTe$_2$ originating from the gate-voltage dependence of the mobility of bulk electrons and holes, caused by surface scattering. Our observations demonstrate the possibility to gate-control the electronic properties of a material well inside its interior, over a depth much larger than the electrostatic screening length. This finding can be relevant for the hydrodynamics of ballistic electrons ~\cite{deJong1995} --since the gate tuning the effect of surface scattering may give control over the viscosity of the electron/hole liquid-- which has attracted enormous interest in recent times~\cite{Bandurin2016,Crossno2016,Moll2016}.

We acknowledge A. Ferreira for technical help. Financial support from the Swiss National Science Foundation and  the EU Graphene Flagship project is also acknowledged.


\begin{thebibliography}{58}%
\makeatletter
\providecommand \@ifxundefined [1]{%
 \@ifx{#1\undefined}
}%
\providecommand \@ifnum [1]{%
 \ifnum #1\expandafter \@firstoftwo
 \else \expandafter \@secondoftwo
 \fi
}%
\providecommand \@ifx [1]{%
 \ifx #1\expandafter \@firstoftwo
 \else \expandafter \@secondoftwo
 \fi
}%
\providecommand \natexlab [1]{#1}%
\providecommand \enquote  [1]{``#1''}%
\providecommand \bibnamefont  [1]{#1}%
\providecommand \bibfnamefont [1]{#1}%
\providecommand \citenamefont [1]{#1}%
\providecommand \href@noop [0]{\@secondoftwo}%
\providecommand \href [0]{\begingroup \@sanitize@url \@href}%
\providecommand \@href[1]{\@@startlink{#1}\@@href}%
\providecommand \@@href[1]{\endgroup#1\@@endlink}%
\providecommand \@sanitize@url [0]{\catcode `\\12\catcode `\$12\catcode
  `\&12\catcode `\#12\catcode `\^12\catcode `\_12\catcode `\%12\relax}%
\providecommand \@@startlink[1]{}%
\providecommand \@@endlink[0]{}%
\providecommand \url  [0]{\begingroup\@sanitize@url \@url }%
\providecommand \@url [1]{\endgroup\@href {#1}{\urlprefix }}%
\providecommand \urlprefix  [0]{URL }%
\providecommand \Eprint [0]{\href }%
\providecommand \doibase [0]{http://dx.doi.org/}%
\providecommand \selectlanguage [0]{\@gobble}%
\providecommand \bibinfo  [0]{\@secondoftwo}%
\providecommand \bibfield  [0]{\@secondoftwo}%
\providecommand \translation [1]{[#1]}%
\providecommand \BibitemOpen [0]{}%
\providecommand \bibitemStop [0]{}%
\providecommand \bibitemNoStop [0]{.\EOS\space}%
\providecommand \EOS [0]{\spacefactor3000\relax}%
\providecommand \BibitemShut  [1]{\csname bibitem#1\endcsname}%
\let\auto@bib@innerbib\@empty
\bibitem [{\citenamefont {Ahn}\ \emph {et~al.}(2006)\citenamefont {Ahn} \emph
  {et~al.}}]{Ahn2006}%
  \BibitemOpen
  \bibfield  {author} {\bibinfo {author} {\bibfnamefont {C.~H.}\ \bibnamefont
  {Ahn}} \emph {et~al.},\ }\href {\doibase 10.1103/RevModPhys.78.1185}
  {\bibfield  {journal} {\bibinfo  {journal} {Rev. Mod. Phys.}\ }\textbf
  {\bibinfo {volume} {78}},\ \bibinfo {pages} {1185} (\bibinfo {year}
  {2006})}\BibitemShut {NoStop}%
\bibitem [{\citenamefont {Ohno}\ \emph {et~al.}(2000)\citenamefont {Ohno} \emph
  {et~al.}}]{Ohno2000}%
  \BibitemOpen
  \bibfield  {author} {\bibinfo {author} {\bibfnamefont {H.}~\bibnamefont
  {Ohno}} \emph {et~al.},\ }\href {\doibase 10.1038/35050040} {\bibfield
  {journal} {\bibinfo  {journal} {Nature (London)}\ }\textbf {\bibinfo {volume}
  {408}},\ \bibinfo {pages} {944} (\bibinfo {year} {2000})}\BibitemShut
  {NoStop}%
\bibitem [{\citenamefont {Novoselov}\ \emph {et~al.}(2004)\citenamefont
  {Novoselov} \emph {et~al.}}]{Novoselov2004}%
  \BibitemOpen
  \bibfield  {author} {\bibinfo {author} {\bibfnamefont {K.~S.}\ \bibnamefont
  {Novoselov}} \emph {et~al.},\ }\href {\doibase 10.1126/science.1102896}
  {\bibfield  {journal} {\bibinfo  {journal} {Science}\ }\textbf {\bibinfo
  {volume} {306}},\ \bibinfo {pages} {666} (\bibinfo {year}
  {2004})}\BibitemShut {NoStop}%
\bibitem [{\citenamefont {Martel}\ \emph {et~al.}(1998)\citenamefont {Martel},
  \citenamefont {Schmidt}, \citenamefont {Shea}, \citenamefont {Hertel},\ and\
  \citenamefont {Avouris}}]{Martel1998}%
  \BibitemOpen
  \bibfield  {author} {\bibinfo {author} {\bibfnamefont {R.}~\bibnamefont
  {Martel}}, \bibinfo {author} {\bibfnamefont {T.}~\bibnamefont {Schmidt}},
  \bibinfo {author} {\bibfnamefont {H.~R.}\ \bibnamefont {Shea}}, \bibinfo
  {author} {\bibfnamefont {T.}~\bibnamefont {Hertel}}, \ and\ \bibinfo {author}
  {\bibfnamefont {P.}~\bibnamefont {Avouris}},\ }\href {\doibase
  10.1063/1.122477} {\bibfield  {journal} {\bibinfo  {journal} {Appl. Phys.
  Lett.}\ }\textbf {\bibinfo {volume} {73}},\ \bibinfo {pages} {2447} (\bibinfo
  {year} {1998})}\BibitemShut {NoStop}%
\bibitem [{\citenamefont {Caviglia}\ \emph {et~al.}(2008)\citenamefont
  {Caviglia} \emph {et~al.}}]{Caviglia2008}%
  \BibitemOpen
  \bibfield  {author} {\bibinfo {author} {\bibfnamefont {A.~D.}\ \bibnamefont
  {Caviglia}} \emph {et~al.},\ }\href {\doibase 10.1038/nature07576} {\bibfield
   {journal} {\bibinfo  {journal} {Nature (London)}\ }\textbf {\bibinfo
  {volume} {456}},\ \bibinfo {pages} {624} (\bibinfo {year}
  {2008})}\BibitemShut {NoStop}%
\bibitem [{\citenamefont {Ali}\ \emph {et~al.}(2014)\citenamefont {Ali} \emph
  {et~al.}}]{Ali2014}%
  \BibitemOpen
  \bibfield  {author} {\bibinfo {author} {\bibfnamefont {M.~N.}\ \bibnamefont
  {Ali}} \emph {et~al.},\ }\href {\doibase 10.1038/nature13763} {\bibfield
  {journal} {\bibinfo  {journal} {Nature (London)}\ }\textbf {\bibinfo {volume}
  {514}},\ \bibinfo {pages} {205} (\bibinfo {year} {2014})}\BibitemShut
  {NoStop}%
\bibitem [{\citenamefont {Pletikosi\'{c}}\ \emph {et~al.}(2014)\citenamefont
  {Pletikosi\'{c}}, \citenamefont {Ali}, \citenamefont {Fedorov}, \citenamefont
  {Cava},\ and\ \citenamefont {Valla}}]{Pletikosic2014}%
  \BibitemOpen
  \bibfield  {author} {\bibinfo {author} {\bibfnamefont {I.}~\bibnamefont
  {Pletikosi\'{c}}}, \bibinfo {author} {\bibfnamefont {M.~N.}\ \bibnamefont
  {Ali}}, \bibinfo {author} {\bibfnamefont {A.~V.}\ \bibnamefont {Fedorov}},
  \bibinfo {author} {\bibfnamefont {R.~J.}\ \bibnamefont {Cava}}, \ and\
  \bibinfo {author} {\bibfnamefont {T.}~\bibnamefont {Valla}},\ }\href@noop {}
  {\bibfield  {journal} {\bibinfo  {journal} {Phys. Rev. Lett.}\ }\textbf
  {\bibinfo {volume} {113}},\ \bibinfo {pages} {216601} (\bibinfo {year}
  {2014})}\BibitemShut {NoStop}%
\bibitem [{\citenamefont {Jiang}\ \emph {et~al.}(2015)\citenamefont {Jiang}
  \emph {et~al.}}]{Jiang2015}%
  \BibitemOpen
  \bibfield  {author} {\bibinfo {author} {\bibfnamefont {J.}~\bibnamefont
  {Jiang}} \emph {et~al.},\ }\href@noop {} {\bibfield  {journal} {\bibinfo
  {journal} {Phys. Rev. Lett.}\ }\textbf {\bibinfo {volume} {115}},\ \bibinfo
  {pages} {166601} (\bibinfo {year} {2015})}\BibitemShut {NoStop}%
\bibitem [{\citenamefont {Zhu}\ \emph {et~al.}(2015)\citenamefont {Zhu} \emph
  {et~al.}}]{Zhu2015}%
  \BibitemOpen
  \bibfield  {author} {\bibinfo {author} {\bibfnamefont {Z.}~\bibnamefont
  {Zhu}} \emph {et~al.},\ }\href@noop {} {\bibfield  {journal} {\bibinfo
  {journal} {Phys. Rev. Lett.}\ }\textbf {\bibinfo {volume} {114}},\ \bibinfo
  {pages} {176601} (\bibinfo {year} {2015})}\BibitemShut {NoStop}%
\bibitem [{\citenamefont {Rhodes}\ \emph {et~al.}(2015)\citenamefont {Rhodes}
  \emph {et~al.}}]{Rhodes2015}%
  \BibitemOpen
  \bibfield  {author} {\bibinfo {author} {\bibfnamefont {D.}~\bibnamefont
  {Rhodes}} \emph {et~al.},\ }\href@noop {} {\bibfield  {journal} {\bibinfo
  {journal} {Phys. Rev. B}\ }\textbf {\bibinfo {volume} {92}},\ \bibinfo
  {pages} {125152} (\bibinfo {year} {2015})}\BibitemShut {NoStop}%
\bibitem [{\citenamefont {Zhao}\ \emph {et~al.}(2015)\citenamefont {Zhao} \emph
  {et~al.}}]{Zhao2015}%
  \BibitemOpen
  \bibfield  {author} {\bibinfo {author} {\bibfnamefont {Y.}~\bibnamefont
  {Zhao}} \emph {et~al.},\ }\href@noop {} {\bibfield  {journal} {\bibinfo
  {journal} {Phys. Rev. B}\ }\textbf {\bibinfo {volume} {92}},\ \bibinfo
  {pages} {041104} (\bibinfo {year} {2015})}\BibitemShut {NoStop}%
\bibitem [{\citenamefont {Cai}\ \emph {et~al.}(2015)\citenamefont {Cai} \emph
  {et~al.}}]{Cai2015}%
  \BibitemOpen
  \bibfield  {author} {\bibinfo {author} {\bibfnamefont {P.~L.}\ \bibnamefont
  {Cai}} \emph {et~al.},\ }\href@noop {} {\bibfield  {journal} {\bibinfo
  {journal} {Phys. Rev. Lett.}\ }\textbf {\bibinfo {volume} {115}},\ \bibinfo
  {pages} {057202} (\bibinfo {year} {2015})}\BibitemShut {NoStop}%
\bibitem [{\citenamefont {Pan}\ \emph {et~al.}(2015)\citenamefont {Pan} \emph
  {et~al.}}]{Pan2015}%
  \BibitemOpen
  \bibfield  {author} {\bibinfo {author} {\bibfnamefont {X.-C.}\ \bibnamefont
  {Pan}} \emph {et~al.},\ }\href {\doibase 10.1038/ncomms8805} {\bibfield
  {journal} {\bibinfo  {journal} {Nat. Commun.}\ }\textbf {\bibinfo {volume}
  {6}},\ \bibinfo {pages} {7805} (\bibinfo {year} {2015})}\BibitemShut
  {NoStop}%
\bibitem [{\citenamefont {Kang}\ \emph {et~al.}(2015)\citenamefont {Kang} \emph
  {et~al.}}]{Kang2015}%
  \BibitemOpen
  \bibfield  {author} {\bibinfo {author} {\bibfnamefont {D.}~\bibnamefont
  {Kang}} \emph {et~al.},\ }\href {\doibase 10.1038/ncomms8804} {\bibfield
  {journal} {\bibinfo  {journal} {Nat. Commun.}\ }\textbf {\bibinfo {volume}
  {6}},\ \bibinfo {pages} {7804} (\bibinfo {year} {2015})}\BibitemShut
  {NoStop}%
\bibitem [{\citenamefont {Thoutam}\ \emph {et~al.}(2015)\citenamefont {Thoutam}
  \emph {et~al.}}]{Thoutam2015}%
  \BibitemOpen
  \bibfield  {author} {\bibinfo {author} {\bibfnamefont {L.~R.}\ \bibnamefont
  {Thoutam}} \emph {et~al.},\ }\href@noop {} {\bibfield  {journal} {\bibinfo
  {journal} {Phys. Rev. Lett.}\ }\textbf {\bibinfo {volume} {115}},\ \bibinfo
  {pages} {046602} (\bibinfo {year} {2015})}\BibitemShut {NoStop}%
\bibitem [{\citenamefont {Wu}\ \emph {et~al.}(2015)\citenamefont {Wu} \emph
  {et~al.}}]{Wu2015}%
  \BibitemOpen
  \bibfield  {author} {\bibinfo {author} {\bibfnamefont {Y.}~\bibnamefont {Wu}}
  \emph {et~al.},\ }\href@noop {} {\bibfield  {journal} {\bibinfo  {journal}
  {Phys. Rev. Lett.}\ }\textbf {\bibinfo {volume} {115}},\ \bibinfo {pages}
  {166602} (\bibinfo {year} {2015})}\BibitemShut {NoStop}%
\bibitem [{\citenamefont {Wang}\ \emph
  {et~al.}(2015{\natexlab{a}})\citenamefont {Wang} \emph {et~al.}}]{Wang2015b}%
  \BibitemOpen
  \bibfield  {author} {\bibinfo {author} {\bibfnamefont {Y.~L.}\ \bibnamefont
  {Wang}} \emph {et~al.},\ }\href@noop {} {\bibfield  {journal} {\bibinfo
  {journal} {Phys. Rev. B}\ }\textbf {\bibinfo {volume} {92}},\ \bibinfo
  {pages} {180402} (\bibinfo {year} {2015}{\natexlab{a}})}\BibitemShut
  {NoStop}%
\bibitem [{\citenamefont {Dai}\ \emph {et~al.}(2015)\citenamefont {Dai} \emph
  {et~al.}}]{Dai2015}%
  \BibitemOpen
  \bibfield  {author} {\bibinfo {author} {\bibfnamefont {Y.~M.}\ \bibnamefont
  {Dai}} \emph {et~al.},\ }\href@noop {} {\bibfield  {journal} {\bibinfo
  {journal} {Phys. Rev. B}\ }\textbf {\bibinfo {volume} {92}},\ \bibinfo
  {pages} {161104} (\bibinfo {year} {2015})}\BibitemShut {NoStop}%
\bibitem [{\citenamefont {Homes}\ \emph {et~al.}(2015)\citenamefont {Homes},
  \citenamefont {Ali},\ and\ \citenamefont {Cava}}]{Homes2015}%
  \BibitemOpen
  \bibfield  {author} {\bibinfo {author} {\bibfnamefont {C.~C.}\ \bibnamefont
  {Homes}}, \bibinfo {author} {\bibfnamefont {M.~N.}\ \bibnamefont {Ali}}, \
  and\ \bibinfo {author} {\bibfnamefont {R.~J.}\ \bibnamefont {Cava}},\
  }\href@noop {} {\bibfield  {journal} {\bibinfo  {journal} {Phys. Rev. B}\
  }\textbf {\bibinfo {volume} {92}},\ \bibinfo {pages} {161109} (\bibinfo
  {year} {2015})}\BibitemShut {NoStop}%
\bibitem [{\citenamefont {Kong}\ \emph {et~al.}(2015)\citenamefont {Kong} \emph
  {et~al.}}]{Kong2015}%
  \BibitemOpen
  \bibfield  {author} {\bibinfo {author} {\bibfnamefont {W.-D.}\ \bibnamefont
  {Kong}} \emph {et~al.},\ }\href {\doibase
  doi:http://dx.doi.org/10.1063/1.4913680} {\bibfield  {journal} {\bibinfo
  {journal} {Appl. Phys. Lett.}\ }\textbf {\bibinfo {volume} {106}},\ \bibinfo
  {pages} {081906} (\bibinfo {year} {2015})}\BibitemShut {NoStop}%
\bibitem [{\citenamefont {Das}\ \emph {et~al.}(2016)\citenamefont {Das} \emph
  {et~al.}}]{Das2016}%
  \BibitemOpen
  \bibfield  {author} {\bibinfo {author} {\bibfnamefont {P.~K.}\ \bibnamefont
  {Das}} \emph {et~al.},\ }\href {\doibase 10.1038/ncomms10847} {\bibfield
  {journal} {\bibinfo  {journal} {Nat. Commun.}\ }\textbf {\bibinfo {volume}
  {7}},\ \bibinfo {pages} {10847} (\bibinfo {year} {2016})}\BibitemShut
  {NoStop}%
\bibitem [{\citenamefont {Wang}\ \emph
  {et~al.}(2015{\natexlab{b}})\citenamefont {Wang}, \citenamefont
  {Guti\'{e}rrez-Lezama}, \citenamefont {Barreteau}, \citenamefont {Ubrig},
  \citenamefont {Giannini},\ and\ \citenamefont {Morpurgo}}]{Wang2015}%
  \BibitemOpen
  \bibfield  {author} {\bibinfo {author} {\bibfnamefont {L.}~\bibnamefont
  {Wang}}, \bibinfo {author} {\bibfnamefont {I.}~\bibnamefont
  {Guti\'{e}rrez-Lezama}}, \bibinfo {author} {\bibfnamefont {C.}~\bibnamefont
  {Barreteau}}, \bibinfo {author} {\bibfnamefont {N.}~\bibnamefont {Ubrig}},
  \bibinfo {author} {\bibfnamefont {E.}~\bibnamefont {Giannini}}, \ and\
  \bibinfo {author} {\bibfnamefont {A.~F.}\ \bibnamefont {Morpurgo}},\ }\href
  {\doibase 10.1038/ncomms9892} {\bibfield  {journal} {\bibinfo  {journal}
  {Nat. Commun.}\ }\textbf {\bibinfo {volume} {6}},\ \bibinfo {pages} {8892}
  (\bibinfo {year} {2015}{\natexlab{b}})}\BibitemShut {NoStop}%
\bibitem [{\citenamefont {Alekseev}\ \emph {et~al.}(2015)\citenamefont
  {Alekseev} \emph {et~al.}}]{Alekseev2015}%
  \BibitemOpen
  \bibfield  {author} {\bibinfo {author} {\bibfnamefont {P.~S.}\ \bibnamefont
  {Alekseev}} \emph {et~al.},\ }\href@noop {} {\bibfield  {journal} {\bibinfo
  {journal} {Phys. Rev. Lett.}\ }\textbf {\bibinfo {volume} {114}},\ \bibinfo
  {pages} {156601} (\bibinfo {year} {2015})}\BibitemShut {NoStop}%
\bibitem [{\citenamefont {Soluyanov}\ \emph {et~al.}(2015)\citenamefont
  {Soluyanov} \emph {et~al.}}]{Soluyanov2015}%
  \BibitemOpen
  \bibfield  {author} {\bibinfo {author} {\bibfnamefont {A.~A.}\ \bibnamefont
  {Soluyanov}} \emph {et~al.},\ }\href {\doibase 10.1038/nature15768}
  {\bibfield  {journal} {\bibinfo  {journal} {Nature (London)}\ }\textbf
  {\bibinfo {volume} {527}},\ \bibinfo {pages} {495} (\bibinfo {year}
  {2015})}\BibitemShut {NoStop}%
\bibitem [{\citenamefont {Qian}\ \emph {et~al.}(2014)\citenamefont {Qian},
  \citenamefont {Liu}, \citenamefont {Fu},\ and\ \citenamefont
  {Li}}]{Qian2014}%
  \BibitemOpen
  \bibfield  {author} {\bibinfo {author} {\bibfnamefont {X.}~\bibnamefont
  {Qian}}, \bibinfo {author} {\bibfnamefont {J.}~\bibnamefont {Liu}}, \bibinfo
  {author} {\bibfnamefont {L.}~\bibnamefont {Fu}}, \ and\ \bibinfo {author}
  {\bibfnamefont {J.}~\bibnamefont {Li}},\ }\href {\doibase
  10.1126/science.1256815} {\bibfield  {journal} {\bibinfo  {journal}
  {Science}\ }\textbf {\bibinfo {volume} {346}},\ \bibinfo {pages} {1344}
  (\bibinfo {year} {2014})}\BibitemShut {NoStop}%
\bibitem [{\citenamefont {Sondheimer}(1952)}]{Sondheimer1952}%
  \BibitemOpen
  \bibfield  {author} {\bibinfo {author} {\bibfnamefont {E.~H.}\ \bibnamefont
  {Sondheimer}},\ }\href@noop {} {\bibfield  {journal} {\bibinfo  {journal}
  {Adv. Phys.}\ }\textbf {\bibinfo {volume} {1}},\ \bibinfo {pages} {1}
  (\bibinfo {year} {1952})}\BibitemShut {NoStop}%
\bibitem [{\citenamefont {Schrieffer}(1955)}]{Schrieffer1955}%
  \BibitemOpen
  \bibfield  {author} {\bibinfo {author} {\bibfnamefont {J.~R.}\ \bibnamefont
  {Schrieffer}},\ }\href@noop {} {\bibfield  {journal} {\bibinfo  {journal}
  {Phys. Rev.}\ }\textbf {\bibinfo {volume} {97}},\ \bibinfo {pages} {641}
  (\bibinfo {year} {1955})}\BibitemShut {NoStop}%
\bibitem [{\citenamefont {Price}(1960)}]{Price1960}%
  \BibitemOpen
  \bibfield  {author} {\bibinfo {author} {\bibfnamefont {P.~J.}\ \bibnamefont
  {Price}},\ }\href {\doibase 10.1147/rd.42.0152} {\bibfield  {journal}
  {\bibinfo  {journal} {IBM J. Res. Dev.}\ }\textbf {\bibinfo {volume} {4}},\
  \bibinfo {pages} {152} (\bibinfo {year} {1960})}\BibitemShut {NoStop}%
\bibitem [{\citenamefont {Ibach}(2006)}]{Ibach2006}%
  \BibitemOpen
  \bibfield  {author} {\bibinfo {author} {\bibfnamefont {H.}~\bibnamefont
  {Ibach}},\ }\href@noop {} {\emph {\bibinfo {title} {Physics of surfaces and
  interfaces}}}\ (\bibinfo  {publisher} {Springer},\ \bibinfo {address} {Berlin
  ; New York},\ \bibinfo {year} {2006})\ pp.\ \bibinfo {pages} {xii, 646
  p.}\BibitemShut {Stop}%
\bibitem [{\citenamefont {Fu}\ \emph {et~al.}(2016{\natexlab{a}})\citenamefont
  {Fu}, \citenamefont {Reich},\ and\ \citenamefont {Shklovskii}}]{Fu2016}%
  \BibitemOpen
  \bibfield  {author} {\bibinfo {author} {\bibfnamefont {H.}~\bibnamefont
  {Fu}}, \bibinfo {author} {\bibfnamefont {K.~V.}\ \bibnamefont {Reich}}, \
  and\ \bibinfo {author} {\bibfnamefont {B.~I.}\ \bibnamefont {Shklovskii}},\
  }\href@noop {} {\bibfield  {journal} {\bibinfo  {journal} {ArXiv e-prints}\ }
  (\bibinfo {year} {2016}{\natexlab{a}})},\ \Eprint
  {http://arxiv.org/abs/1604.08509} {arXiv:1604.08509 [cond-mat.mes-hall]}
  \BibitemShut {NoStop}%
\bibitem [{\citenamefont {Fu}\ \emph {et~al.}(2016{\natexlab{b}})\citenamefont
  {Fu}, \citenamefont {Reich},\ and\ \citenamefont {Shklovskii}}]{Fu2016a}%
  \BibitemOpen
  \bibfield  {author} {\bibinfo {author} {\bibfnamefont {H.}~\bibnamefont
  {Fu}}, \bibinfo {author} {\bibfnamefont {K.~V.}\ \bibnamefont {Reich}}, \
  and\ \bibinfo {author} {\bibfnamefont {B.~I.}\ \bibnamefont {Shklovskii}},\
  }\href@noop {} {\bibfield  {journal} {\bibinfo  {journal} {ArXiv e-prints}\ }
  (\bibinfo {year} {2016}{\natexlab{b}})},\ \Eprint
  {http://arxiv.org/abs/1603.03676} {arXiv:1603.03676 [cond-mat.mes-hall]}
  \BibitemShut {NoStop}%
\bibitem [{Note1()}]{Note1}%
  \BibitemOpen
  \bibinfo {note} {The non locality that we are referring to is described by a
  relation between current density $\protect \mathaccentV {vec}17E{J}$ and
  electrical field $\protect \mathaccentV {vec}17E{E}$, of the type $\protect
  \mathaccentV {vec}17E{J}(x)=\DOTSI \intop \ilimits@ dy\sigma (x,y)\protect
  \mathaccentV {vec}17E{E}(y)$}\BibitemShut {NoStop}%
\bibitem [{\citenamefont {Reuter}\ and\ \citenamefont
  {Sondheimer}(1948)}]{Reuter1948}%
  \BibitemOpen
  \bibfield  {author} {\bibinfo {author} {\bibfnamefont {G.~E.~H.}\
  \bibnamefont {Reuter}}\ and\ \bibinfo {author} {\bibfnamefont {E.~H.}\
  \bibnamefont {Sondheimer}},\ }\href@noop {} {\bibfield  {journal} {\bibinfo
  {journal} {Proc. Roy. Soc. London. Ser. A}\ }\textbf {\bibinfo {volume}
  {195}},\ \bibinfo {pages} {336} (\bibinfo {year} {1948})}\BibitemShut
  {NoStop}%
\bibitem [{\citenamefont {Chambers}(1950)}]{Chambers1950}%
  \BibitemOpen
  \bibfield  {author} {\bibinfo {author} {\bibfnamefont {R.~G.}\ \bibnamefont
  {Chambers}},\ }\href@noop {} {\bibfield  {journal} {\bibinfo  {journal}
  {Nature (London)}\ }\textbf {\bibinfo {volume} {165}},\ \bibinfo {pages}
  {239} (\bibinfo {year} {1950})}\BibitemShut {NoStop}%
\bibitem [{\citenamefont {Chambers}(1952)}]{Chambers1952}%
  \BibitemOpen
  \bibfield  {author} {\bibinfo {author} {\bibfnamefont {R.~G.}\ \bibnamefont
  {Chambers}},\ }\href@noop {} {\bibfield  {journal} {\bibinfo  {journal}
  {Proc. Roy. Soc. London. Ser. A}\ }\textbf {\bibinfo {volume} {215}},\
  \bibinfo {pages} {481} (\bibinfo {year} {1952})}\BibitemShut {NoStop}%
\bibitem [{\citenamefont {Berman}\ and\ \citenamefont
  {Juretschke}(1971)}]{Berman1971}%
  \BibitemOpen
  \bibfield  {author} {\bibinfo {author} {\bibfnamefont {A.}~\bibnamefont
  {Berman}}\ and\ \bibinfo {author} {\bibfnamefont {H.}~\bibnamefont
  {Juretschke}},\ }\href@noop {} {\bibfield  {journal} {\bibinfo  {journal}
  {Appl. Phys. Lett.}\ }\textbf {\bibinfo {volume} {18}},\ \bibinfo {pages}
  {417} (\bibinfo {year} {1971})}\BibitemShut {NoStop}%
\bibitem [{\citenamefont {Berman}\ and\ \citenamefont
  {Juretschke}(1975)}]{Berman1975}%
  \BibitemOpen
  \bibfield  {author} {\bibinfo {author} {\bibfnamefont {A.}~\bibnamefont
  {Berman}}\ and\ \bibinfo {author} {\bibfnamefont {H.~J.}\ \bibnamefont
  {Juretschke}},\ }\href@noop {} {\bibfield  {journal} {\bibinfo  {journal}
  {Phys. Rev. B}\ }\textbf {\bibinfo {volume} {11}},\ \bibinfo {pages} {2903}
  (\bibinfo {year} {1975})}\BibitemShut {NoStop}%
\bibitem [{Note2()}]{Note2}%
  \BibitemOpen
  \bibinfo {note} {In Refs.~\cite {Berman1971,Berman1975} the observation of a
  (gate-induced) 10$^{-5}$ relative change in the conductivity of metallic thin
  films was claimed to represent a manifestation of these phenomena. However,
  no actual evidence for a long-range field-effect was presented, as the claim
  was entirely based on data interpretation (using the phenomenological
  Fuchs-Sondheimer model) and not on direct measurements. It is now known that
  the assumptions made in Refs.~\cite {Berman1971,Berman1975} are wrong and so
  are the conclusions (thorough discussions of this issue can be found in
  numerous papers, including A. F. Mayadas \protect \textit {et al.}, Appl.
  Phys. Lett. \protect \textbf {14}, 345 (1969); A. F. Mayadas and M. Shatzkes,
  Phys. Rev. B \protect \textbf {11}, 1382 (1970); R. C. Munoz \protect \textit
  {et al.}, Phys. Rev. B \protect \textbf {62}, 7 (2000))}\BibitemShut
  {NoStop}%
\bibitem [{Note3()}]{Note3}%
  \BibitemOpen
  \bibinfo {note} {For Sample A, for instance, $\mu _e=2.900$
  cm$^2$V$^{-1}$s$^{-1}$ and $\mu _h=2.900$ cm$^2$V$^{-1}$s$^{-1}$, so that the
  electron and hole mean-free paths $L_h\sim L_e\sim 0.3$ $\mu $m $\gg t$; the
  Fermi vector $k_F$ is estimated from Refs.~\cite
  {Pletikosic2014,Jiang2015,Rhodes2015,Zhu2015}}\BibitemShut {NoStop}%
\bibitem [{sup()}]{support}%
  \BibitemOpen
  \href@noop {} {\ }\bibinfo {note} {See Supplemental Material at [URL will be
  inserted by publisher] for the details of the quantitative analysis of
  classical magneto-transport behavior and the gate-dependent Shubnikov de Hass
  oscillations in WTe$_2$}\BibitemShut {NoStop}%
\bibitem [{\citenamefont {Murzin}\ \emph {et~al.}(1946)\citenamefont {Murzin},
  \citenamefont {Dorozhkin}, \citenamefont {Landwehr},\ and\ \citenamefont
  {Gossard}}]{Murzin1946}%
  \BibitemOpen
  \bibfield  {author} {\bibinfo {author} {\bibfnamefont {S.~S.}\ \bibnamefont
  {Murzin}}, \bibinfo {author} {\bibfnamefont {S.~I.}\ \bibnamefont
  {Dorozhkin}}, \bibinfo {author} {\bibfnamefont {G.}~\bibnamefont {Landwehr}},
  \ and\ \bibinfo {author} {\bibfnamefont {A.~C.}\ \bibnamefont {Gossard}},\
  }\href {\doibase 10.1134/1.567643} {\bibfield  {journal} {\bibinfo  {journal}
  {JETP Lett.}\ }\textbf {\bibinfo {volume} {67}},\ \bibinfo {pages} {113}
  (\bibinfo {year} {1946})}\BibitemShut {NoStop}%
\bibitem [{\citenamefont {Sondheimer}\ and\ \citenamefont
  {Wilson}(1947)}]{Sondheimer1947}%
  \BibitemOpen
  \bibfield  {author} {\bibinfo {author} {\bibfnamefont {E.~H.}\ \bibnamefont
  {Sondheimer}}\ and\ \bibinfo {author} {\bibfnamefont {A.~H.}\ \bibnamefont
  {Wilson}},\ }\href@noop {} {\bibfield  {journal} {\bibinfo  {journal} {Proc.
  Roy. Soc. London. Ser. A}\ }\textbf {\bibinfo {volume} {190}},\ \bibinfo
  {pages} {435} (\bibinfo {year} {1947})}\BibitemShut {NoStop}%
\bibitem [{Note4()}]{Note4}%
  \BibitemOpen
  \bibinfo {note} {This conclusion is confirmed by the observation, discussed
  in the Supplemental Material~\cite{support}, that when the exposure to
  ambient of the WTe$_2$ crystals is only very limited --so that only minor
  surface degradation occurs-- no gate-voltage dependence of transport is
  observed experimentally. In general the magnitude of the field-effect that we
  observe depends on the amount of surface degradation which is determined by
  details of the fabrication process.}\BibitemShut {Stop}%
\bibitem [{\citenamefont {Zhang}\ \emph
  {et~al.}(2005{\natexlab{a}})\citenamefont {Zhang}, \citenamefont {Small},
  \citenamefont {Amori},\ and\ \citenamefont {Kim}}]{Zhang2005}%
  \BibitemOpen
  \bibfield  {author} {\bibinfo {author} {\bibfnamefont {Y.}~\bibnamefont
  {Zhang}}, \bibinfo {author} {\bibfnamefont {J.~P.}\ \bibnamefont {Small}},
  \bibinfo {author} {\bibfnamefont {M.~E.~S.}\ \bibnamefont {Amori}}, \ and\
  \bibinfo {author} {\bibfnamefont {P.}~\bibnamefont {Kim}},\ }\href@noop {}
  {\bibfield  {journal} {\bibinfo  {journal} {Phys. Rev. Lett.}\ }\textbf
  {\bibinfo {volume} {94}},\ \bibinfo {pages} {176803} (\bibinfo {year}
  {2005}{\natexlab{a}})}\BibitemShut {NoStop}%
\bibitem [{\citenamefont {Zhang}\ \emph
  {et~al.}(2005{\natexlab{b}})\citenamefont {Zhang}, \citenamefont {Small},
  \citenamefont {Pontius},\ and\ \citenamefont {Kim}}]{Zhang2005a}%
  \BibitemOpen
  \bibfield  {author} {\bibinfo {author} {\bibfnamefont {Y.}~\bibnamefont
  {Zhang}}, \bibinfo {author} {\bibfnamefont {J.~P.}\ \bibnamefont {Small}},
  \bibinfo {author} {\bibfnamefont {W.~V.}\ \bibnamefont {Pontius}}, \ and\
  \bibinfo {author} {\bibfnamefont {P.}~\bibnamefont {Kim}},\ }\href {\doibase
  doi:http://dx.doi.org/10.1063/1.1862334} {\bibfield  {journal} {\bibinfo
  {journal} {Appl. Phys. Lett.}\ }\textbf {\bibinfo {volume} {86}},\ \bibinfo
  {pages} {073104} (\bibinfo {year} {2005}{\natexlab{b}})}\BibitemShut
  {NoStop}%
\bibitem [{\citenamefont {Butenko}\ \emph {et~al.}(1997)\citenamefont
  {Butenko}, \citenamefont {Sandomirsky}, \citenamefont {Schlesinger},
  \citenamefont {Shvarts},\ and\ \citenamefont {Sokol}}]{Butenko1997}%
  \BibitemOpen
  \bibfield  {author} {\bibinfo {author} {\bibfnamefont {A.~V.}\ \bibnamefont
  {Butenko}}, \bibinfo {author} {\bibfnamefont {V.}~\bibnamefont
  {Sandomirsky}}, \bibinfo {author} {\bibfnamefont {Y.}~\bibnamefont
  {Schlesinger}}, \bibinfo {author} {\bibfnamefont {D.}~\bibnamefont
  {Shvarts}}, \ and\ \bibinfo {author} {\bibfnamefont {V.~A.}\ \bibnamefont
  {Sokol}},\ }\href {\doibase doi:http://dx.doi.org/10.1063/1.365897}
  {\bibfield  {journal} {\bibinfo  {journal} {J. Appl. Phys.}\ }\textbf
  {\bibinfo {volume} {82}},\ \bibinfo {pages} {1266} (\bibinfo {year}
  {1997})}\BibitemShut {NoStop}%
\bibitem [{\citenamefont {Butenko}\ \emph {et~al.}(1999)\citenamefont
  {Butenko}, \citenamefont {Shvarts}, \citenamefont {Sandomirsky},\ and\
  \citenamefont {Schlesinger}}]{Butenko1999}%
  \BibitemOpen
  \bibfield  {author} {\bibinfo {author} {\bibfnamefont {A.~V.}\ \bibnamefont
  {Butenko}}, \bibinfo {author} {\bibfnamefont {D.}~\bibnamefont {Shvarts}},
  \bibinfo {author} {\bibfnamefont {V.}~\bibnamefont {Sandomirsky}}, \ and\
  \bibinfo {author} {\bibfnamefont {Y.}~\bibnamefont {Schlesinger}},\ }\href
  {\doibase doi:http://dx.doi.org/10.1063/1.124776} {\bibfield  {journal}
  {\bibinfo  {journal} {Appl. Phys. Lett.}\ }\textbf {\bibinfo {volume} {75}},\
  \bibinfo {pages} {1628} (\bibinfo {year} {1999})}\BibitemShut {NoStop}%
\bibitem [{\citenamefont {Ando}\ \emph {et~al.}(1982)\citenamefont {Ando},
  \citenamefont {Fowler},\ and\ \citenamefont {Stern}}]{Ando1982}%
  \BibitemOpen
  \bibfield  {author} {\bibinfo {author} {\bibfnamefont {T.}~\bibnamefont
  {Ando}}, \bibinfo {author} {\bibfnamefont {A.~B.}\ \bibnamefont {Fowler}}, \
  and\ \bibinfo {author} {\bibfnamefont {F.}~\bibnamefont {Stern}},\
  }\href@noop {} {\bibfield  {journal} {\bibinfo  {journal} {Rev. Mod. Phys.}\
  }\textbf {\bibinfo {volume} {54}},\ \bibinfo {pages} {437} (\bibinfo {year}
  {1982})}\BibitemShut {NoStop}%
\bibitem [{\citenamefont {Richards}(1973)}]{Richards1973}%
  \BibitemOpen
  \bibfield  {author} {\bibinfo {author} {\bibfnamefont {F.~E.}\ \bibnamefont
  {Richards}},\ }\href@noop {} {\bibfield  {journal} {\bibinfo  {journal}
  {Phys. Rev. B}\ }\textbf {\bibinfo {volume} {8}},\ \bibinfo {pages} {2552}
  (\bibinfo {year} {1973})}\BibitemShut {NoStop}%
\bibitem [{\citenamefont {Niederer}(1974)}]{Niederer1974}%
  \BibitemOpen
  \bibfield  {author} {\bibinfo {author} {\bibfnamefont {H.~H. J.~M.}\
  \bibnamefont {Niederer}},\ }\href@noop {} {\bibfield  {journal} {\bibinfo
  {journal} {JPN. J. Appl. Phys.}\ }\textbf {\bibinfo {volume} {13}},\ \bibinfo
  {pages} {339} (\bibinfo {year} {1974})}\BibitemShut {NoStop}%
\bibitem [{\citenamefont {Eisele}\ \emph {et~al.}(1976)\citenamefont {Eisele},
  \citenamefont {Gesch},\ and\ \citenamefont {Dorda}}]{Eisele1976}%
  \BibitemOpen
  \bibfield  {author} {\bibinfo {author} {\bibfnamefont {I.}~\bibnamefont
  {Eisele}}, \bibinfo {author} {\bibfnamefont {H.}~\bibnamefont {Gesch}}, \
  and\ \bibinfo {author} {\bibfnamefont {G.}~\bibnamefont {Dorda}},\ }\href
  {\doibase http://dx.doi.org/10.1016/0039-6028(76)90130-8} {\bibfield
  {journal} {\bibinfo  {journal} {Surf. Sci.}\ }\textbf {\bibinfo {volume}
  {58}},\ \bibinfo {pages} {169} (\bibinfo {year} {1976})}\BibitemShut
  {NoStop}%
\bibitem [{\citenamefont {Fang}\ \emph {et~al.}(1977)\citenamefont {Fang},
  \citenamefont {Fowler},\ and\ \citenamefont {Hartstein}}]{Fang1977}%
  \BibitemOpen
  \bibfield  {author} {\bibinfo {author} {\bibfnamefont {F.~F.}\ \bibnamefont
  {Fang}}, \bibinfo {author} {\bibfnamefont {A.~B.}\ \bibnamefont {Fowler}}, \
  and\ \bibinfo {author} {\bibfnamefont {A.}~\bibnamefont {Hartstein}},\
  }\href@noop {} {\bibfield  {journal} {\bibinfo  {journal} {Phys. Rev. B}\
  }\textbf {\bibinfo {volume} {16}},\ \bibinfo {pages} {4446} (\bibinfo {year}
  {1977})}\BibitemShut {NoStop}%
\bibitem [{\citenamefont {Bangura}\ \emph {et~al.}(2008)\citenamefont {Bangura}
  \emph {et~al.}}]{Bangura2008}%
  \BibitemOpen
  \bibfield  {author} {\bibinfo {author} {\bibfnamefont {A.~F.}\ \bibnamefont
  {Bangura}} \emph {et~al.},\ }\href@noop {} {\bibfield  {journal} {\bibinfo
  {journal} {Phys. Rev. Lett.}\ }\textbf {\bibinfo {volume} {100}},\ \bibinfo
  {pages} {047004} (\bibinfo {year} {2008})}\BibitemShut {NoStop}%
\bibitem [{Note5()}]{Note5}%
  \BibitemOpen
  \bibinfo {note} {For the electron/hole effective masses we use
  $m_e^*=0.5m_0$, $m_h^*=1.0m_0$, in the range of values reported in the
  literature~\cite {Pletikosic2014,Jiang2015,Rhodes2015,Zhu2015}; $m_e^*=
  0.33\sim 0.51 m_0$ and $m_h^*= 0.42\sim 1.1 m_0$, with $m_0$ the free
  electron mass; for $f$, we insert the observed value}\BibitemShut {NoStop}%
\bibitem [{\citenamefont {de~Jong}\ and\ \citenamefont
  {Molenkamp}(1995)}]{deJong1995}%
  \BibitemOpen
  \bibfield  {author} {\bibinfo {author} {\bibfnamefont {M.~J.~M.}\
  \bibnamefont {de~Jong}}\ and\ \bibinfo {author} {\bibfnamefont {L.~W.}\
  \bibnamefont {Molenkamp}},\ }\href@noop {} {\bibfield  {journal} {\bibinfo
  {journal} {Phys. Rev. B}\ }\textbf {\bibinfo {volume} {51}},\ \bibinfo
  {pages} {13389} (\bibinfo {year} {1995})}\BibitemShut {NoStop}%
\bibitem [{\citenamefont {Bandurin}\ \emph {et~al.}(2016)\citenamefont
  {Bandurin} \emph {et~al.}}]{Bandurin2016}%
  \BibitemOpen
  \bibfield  {author} {\bibinfo {author} {\bibfnamefont {D.~A.}\ \bibnamefont
  {Bandurin}} \emph {et~al.},\ }\href {\doibase 10.1126/science.aad0201}
  {\bibfield  {journal} {\bibinfo  {journal} {Science}\ }\textbf {\bibinfo
  {volume} {351}},\ \bibinfo {pages} {1055} (\bibinfo {year}
  {2016})}\BibitemShut {NoStop}%
\bibitem [{\citenamefont {Crossno}\ \emph {et~al.}(2016)\citenamefont {Crossno}
  \emph {et~al.}}]{Crossno2016}%
  \BibitemOpen
  \bibfield  {author} {\bibinfo {author} {\bibfnamefont {J.}~\bibnamefont
  {Crossno}} \emph {et~al.},\ }\href {\doibase 10.1126/science.aad0343}
  {\bibfield  {journal} {\bibinfo  {journal} {Science}\ }\textbf {\bibinfo
  {volume} {351}},\ \bibinfo {pages} {1058} (\bibinfo {year}
  {2016})}\BibitemShut {NoStop}%
\bibitem [{\citenamefont {Moll}\ \emph {et~al.}(2016)\citenamefont {Moll},
  \citenamefont {Kushwaha}, \citenamefont {Nandi}, \citenamefont {Schmidt},\
  and\ \citenamefont {Mackenzie}}]{Moll2016}%
  \BibitemOpen
  \bibfield  {author} {\bibinfo {author} {\bibfnamefont {P.~J.~W.}\
  \bibnamefont {Moll}}, \bibinfo {author} {\bibfnamefont {P.}~\bibnamefont
  {Kushwaha}}, \bibinfo {author} {\bibfnamefont {N.}~\bibnamefont {Nandi}},
  \bibinfo {author} {\bibfnamefont {B.}~\bibnamefont {Schmidt}}, \ and\
  \bibinfo {author} {\bibfnamefont {A.~P.}\ \bibnamefont {Mackenzie}},\ }\href
  {\doibase 10.1126/science.aac8385} {\bibfield  {journal} {\bibinfo  {journal}
  {Science}\ }\textbf {\bibinfo {volume} {351}},\ \bibinfo {pages} {1061}
  (\bibinfo {year} {2016})}\BibitemShut {NoStop}%
\end{thebibliography}

\clearpage

\section*{\refname}

\begin{thebibliography}{11}
\bibitem{R2} S.S. Murzin, S.I. Dorozhkin, G. Landwehr, and A.C. Gossard, JETP Lett. \textbf{67}, 113 (1946).
\bibitem{R3} E.H. Sondheimer and A.H. Wilson, Proc. Roy. Soc. London. Ser. A \textbf{190}, 435 (1947).
\end{thebibliography}
%

\pagebreak
\widetext
\begin{center}
\textbf{\large Supplemental Materials}
\end{center}
\setcounter{equation}{0}
\setcounter{figure}{0}
\setcounter{table}{0}
\makeatletter
\renewcommand{\theequation}{S\arabic{equation}}
\renewcommand{\thefigure}{S\arabic{figure}}
\renewcommand{\bibnumfmt}[1]{[S#1]}
\renewcommand{\citenumfont}[1]{S#1}
\renewcommand{\refname}{\large{Supplemental References:}}

The purpose of the material presented here is to show that

1)  The gate-dependent classical and quantum magneto-transport behavior discussed in the main text is reproducible in all devices investigated.

2) the quantitative analysis of the classical magneto-transport also reproduces measurements done on thin WTe$_2$ flakes.

\section{1. Quantitative analysis of gate-dependent classical magneto-transport of $\mathrm{WTe}_2$ crystals}
The gate-dependent magneto-transport behavior discussed and analyzed in the main text was reproducibly observed in more than 20 devices, having thickness ranging from 10 to 50 nm. Qualitatively identical gate-induced changes of magneto-transport have been observed in crystals with thickness between 100 and 200 nm, in which the magnitude of the gate-induced change of the conductivity was still up to a couple of percent. We did not perform a systematic study of the thickness dependence of the gate-induced changes because the magnitude of the effect depends on the extent of surface degradation, which in turns depends on the history of the fabrication process (and specifically on how long and under which conditions an exfoliated crystal is exposed to air; see Fig. S1). Out of the approximately 20 devices having thickness between 10 and 50 nm, a full quantitative analysis using the two-band model was performed on six of them. 

The exact details of the data analysis differ for thick ($t>15$ nm) and thin ($t\sim10$-15 nm) crystals due to the different size of the contribution from the surface conductivity to the measured total conductivity. For thick crystals (as discussed in the main text), the surface contribution to transport induced by $V_g$ can be entirely neglected (see Fig. S2 for the additional confirmation; Samples D and F), whereas for the crystals with a thickness of 10-15 nm, the gate modulation of the surface contribution cannot be entirely ignored although it is still small compared to the "bulk" one (because of very low carrier mobility on the surface). 

Here we discuss in detail the case for thin WTe$_2$ crystals of the thickness of 10-11 nm: Samples B (11 nm thick), C (11 nm thick) and E (10 nm thick). In this case, the interface longitudinal ($\sigma_{xx,inter}$) and transverse ($\sigma_{xy,inter}$) conductance of the Eqs. (1-2) in the main text should be included in the total conductivity, which reads~\cite{R2,R3}:
\begin{align}
\sigma_{xx,inter} & = \frac{p'e\mu_h'}{1+\left(\mu_h'B\right)^2}+\frac{n'e\mu_e'}{1+\left(\mu_e'B\right)^2} = p'e\mu_h' + n'e\mu_e' = \sigma_{inter}^0 \\ 
\sigma_{xy,inter}  &= \left(\frac{p'e\mu_h'^2}{1+\left(\mu_h'B\right)^2}-\frac{n'e\mu_e'^2}{1+\left(\mu_e'B\right)^2}\right)B = \left(p'e\mu_h'^2-n'e\mu_e'^2\right) = kB.
\end{align}
in terms of the density and mobility of electrons and holes at the interface ($n'$, $p'$, $\mu_e'$ and $\mu_h'$). Since the mobility of electrons and holes is much lower at the interface than in the bulk, $\mu_{e,h}'B<<1$ for all values of $B$ in the experiments, and thus, the denominator in Eqs. (S1-S2) can be neglected. As a result, including the interface contribution to transport adds only two parameters, $\sigma_{inter}^0$ and $k$, which can be determined unambiguously due to the rich behavior of the measured magneto-resistances and in particular of the measured transverse conductivity $\sigma_{xy}$. The precise agreement between theory and data is clearly illustrated in Fig. 3 of the main text for Sample B, and in Fig. S2 for Samples C and E.

We can verify the consistency of our assumptions by looking at the values of the extracted fitting parameters. Fig. S2 shows the interface contribution $\sigma_{inter}^0$ to the longitudinal conductivity extracted from the data analysis of Samples B, C and E. We find, as we expected, that for all gate voltages $\sigma_{inter}^0$ is only a few percent of the total measured longitudinal square conductance of the device. This is the case for all the devices based on 10-11 nm thick crystals that we have investigated in detail, in which the change in the interface contribution to the measured conductance at $B=0$ T with gate voltage is in all cases less than 5 \%. Note that the gate dependence of the surface conductivity correlates with the magnitude of the electron and hole mobility. In particular, for Sample B the conductivity decreases when making $V_g$ more positive, because holes have larger mobility than electrons. Conversely, in Sample C and E the conductivity increases when $V_g$ becomes more positive, consistently with electrons having larger mobility than holes in these devices.

\section{2. Gate-dependent Shubnikov-de Haas oscillations}
In the main text we have discussed the behavior of the Shubnikov-de Haas (SdH) resistance oscillations by showing the data of two devices. Here we present data from more devices --Samples E and G, which are analogous respectively to those of Samples C and D shown in Fig. 4-- to show how the considerations made in the main text are of general validity.

Fig. S4(a) shows the Fourier spectrum of the SdH oscillations of Sample E, which at $V_g=0$ V exhibits a single broad peak at  $f\sim114$ T, with a shoulder at $f\sim144$ T. These features correspond to those observed in Sample C discussed in the main text (see Fig. 4(c); in Sample E the shoulder is considerably more pronounced). As $V_g$ is increased from -80 V to +80 V (which causes a decrease in $\mu_e$ and an increase in $\mu_h$), the height of the main peak is strongly suppressed, and --at the largest positive gate voltage-- the shoulder develops itself into a well-defined peak. The observed features can be reproduced at a semi-quantitative level by using Eq. (5) of the main text, with the values of $\mu_e(V_g)$ and $\mu_h(V_g)$ extracted from the analysis of classical magneto-transport (see Fig. S2(c) for the mobility values). In this device, therefore, varying $V_g$ results in a pronounced change in the strength of the electron and hole contribution to the SdH oscillations, so that at large positive $V_g$ the hole contribution dominates. This is consistent with the fact that at $V_g=+80$ V $\mu_h>\mu_e$. The smaller amplitude of the hole peak amplitude in Sample C and its weaker gate dependence as compared to Sample E are consistent with the lower hole mobility found in Sample C (see Fig. S2(c); note that Sample C corresponds to Sample B, which was intentionally exposed to air for 12 hours). 

The situation is generally different for thicker crystals (40-50 nm), in which the higher mobility very frequently allows the observation of both electron and hole behavior in the oscillation spectrum. Sample G --whose data are shown in Fig. S4(b)-- confirms the behavior shown by Sample D discussed in the main text (Fig. 4(d)), with peaks in the spectrum that exhibit electron behavior (i.e., amplitude decreasing upon increasing $V_g$) in the range 75 T $<f<$ 105 T, and hole behavior (i.e., amplitude increasing upon increasing $V_g$) for 115 T $<f<$ 165 T.


\begin{figure}[b]
\includegraphics[width=13cm]{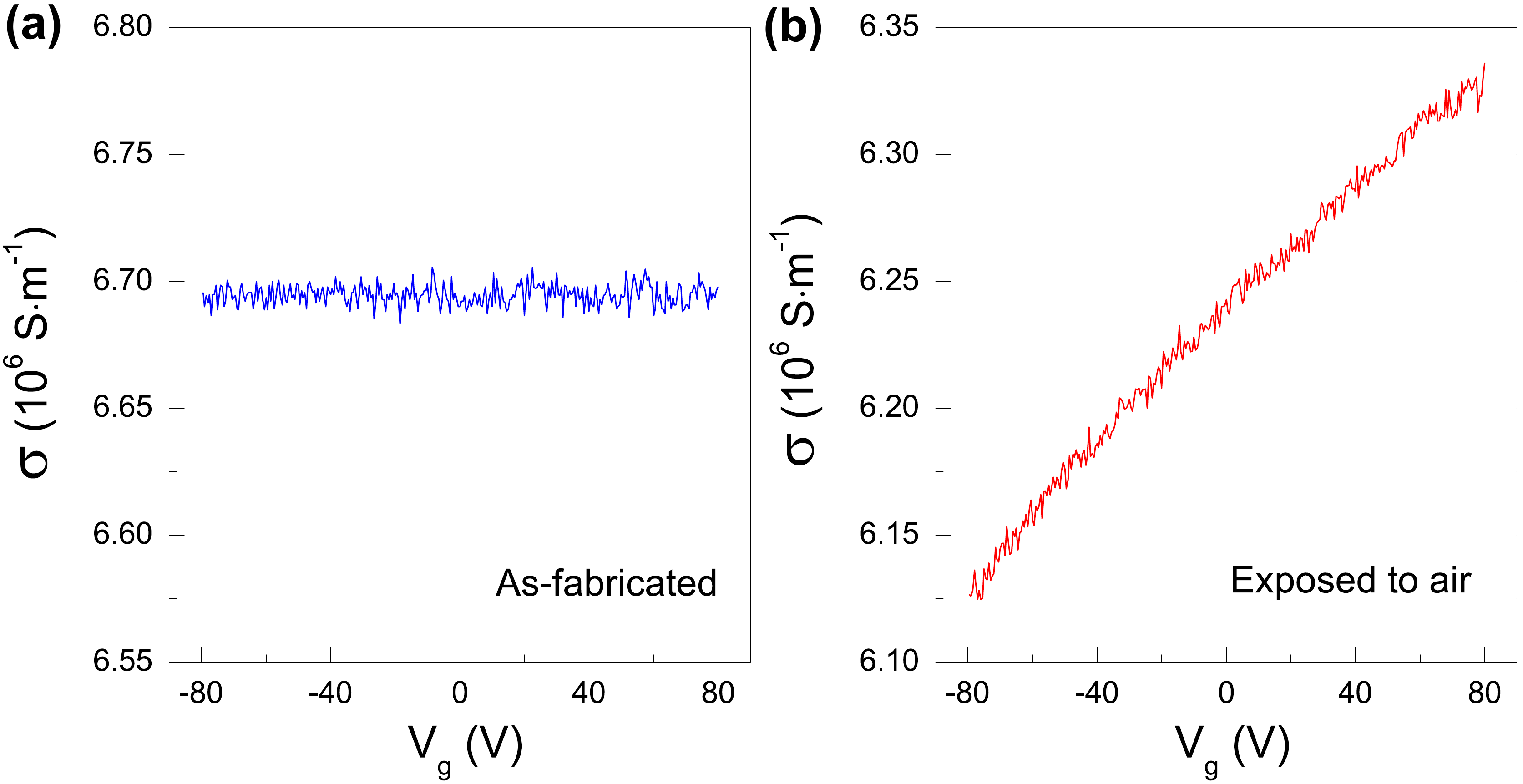}
\caption{\label{figS1} (Color online) \textbf{Influence of surface degradation on the gate-dependence of the measured conductivity.} \textbf{(a)} and \textbf{(b)} show the conductivity $\sigma$ of a 41 nm thick sample as a function of gate voltage $V_g$ before (blue) and after (red) exposure to ambient. Prior to exposure no gate voltage dependence is observed; after exposure a clear gate voltage dependence is seen, because surface degradation –and hence surface scattering– occurs due to exposure to air. In general, the amount of degradation depends on the details of device fabrication, e.g. on whether prolonged exposure to air occurred before transferring the exfoliated crystal onto the Si/SiO$_2$ substrate or after (the larger gate effect is observed when the surface of the exfoliated crystal is exposed to air prior to transferring). Irrespective of these details, thedata shown here clearly confirm that the gate dependence of transport originates from surface degradation. The measurements in the two panels were performed at $T= 250$ mK.}
\end{figure}

\begin{figure}[b]
\includegraphics[width=13cm]{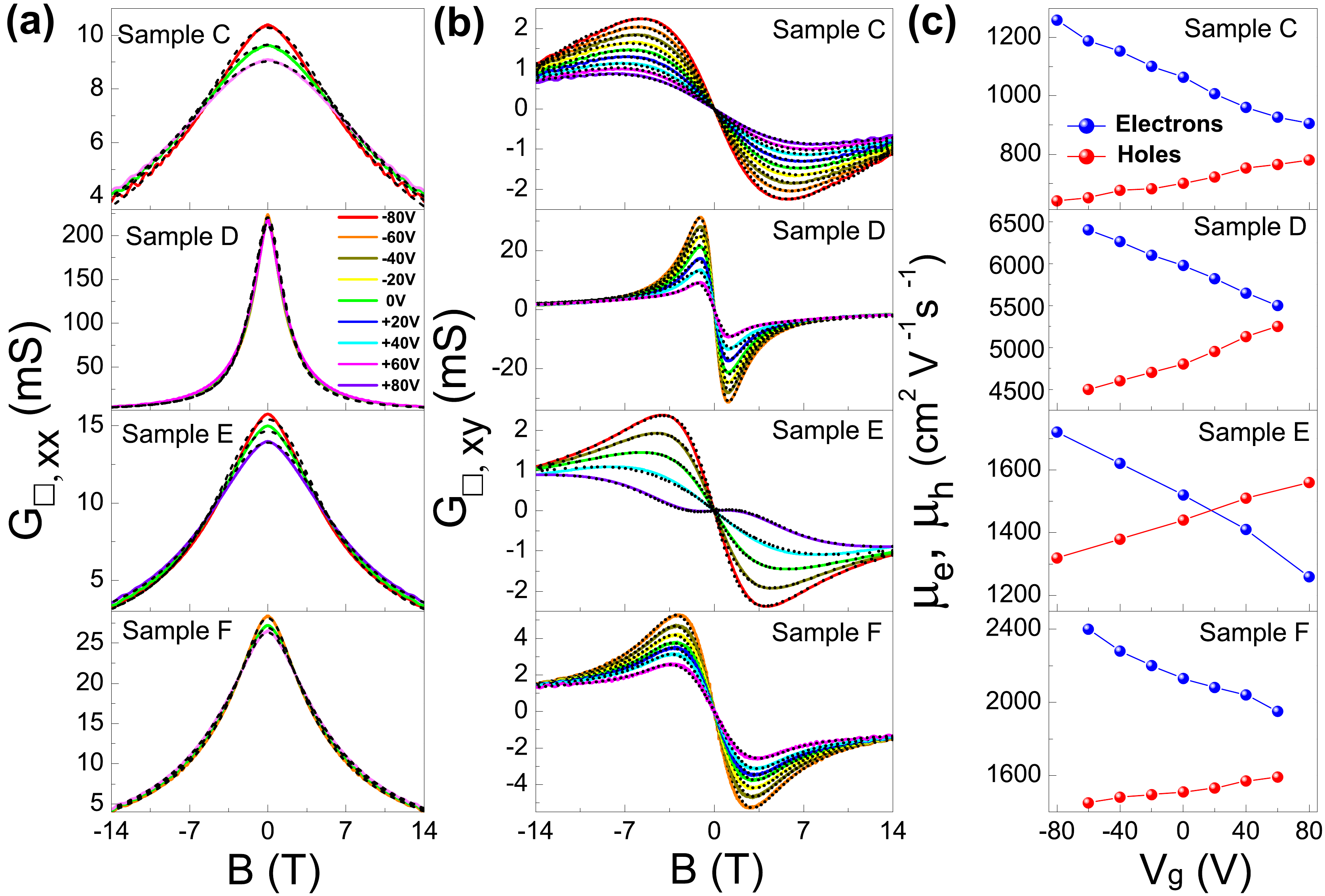}
\caption{\label{figS2} (Color online) \textbf{Quantitative analysis of gate-dependent classical magneto-transport: data from additional devices.} Samples C, D, E and F have been realized using WTe$_2$ crystals that are respectively 11 nm, 37 nm, 10 nm and 16 nm thick. \textbf{(a)} Longitudinal square conductance $G_{\square,xx}$ of the devices. \textbf{(b)} Transverse square conductance $G_{\square,xy}$ of the same devices. The colored curves represent the measured data and the black dotted lines the theoretical fit (in all of these panels, curves of a same color correspond to the same gate voltage value, as indicated by the legend in panel (a); all measurements were taken at $T=250$ mK). \textbf{(c)} Mobility of electrons (blue) and holes (red) extracted from fitting the magneto-transport data of the different devices, as a function of gate voltage $V_g$.}
\end{figure}

\begin{figure}[t]
\includegraphics[width=13cm]{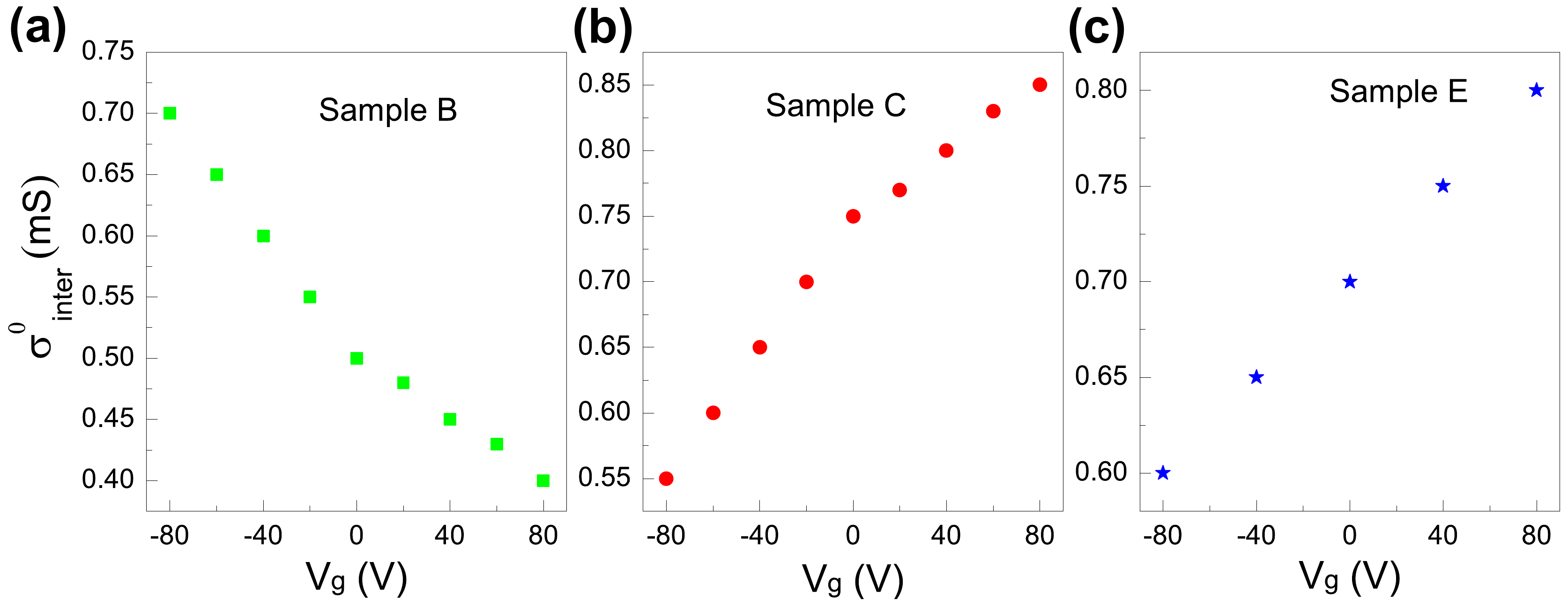}
\caption{\label{figS3} (Color online) \textbf{Gate voltage dependent interface conductivity in thin WTe2 crystals.} \textbf{(a, b, c)} Gate-voltage $V_g$ dependence of the $B=0$ T interface conductivity $\sigma_{inter}^0$ as extracted from fitting the classical magneto-transport properties of Sample B, C, and E, respectively. For all devices investigated, the change in interface contribution to transport $\sigma_{inter}^0$ with gate voltage is at most 5 \% of the total measured square conductance.}
\end{figure}

\begin{figure}[b]
\includegraphics[width=13cm]{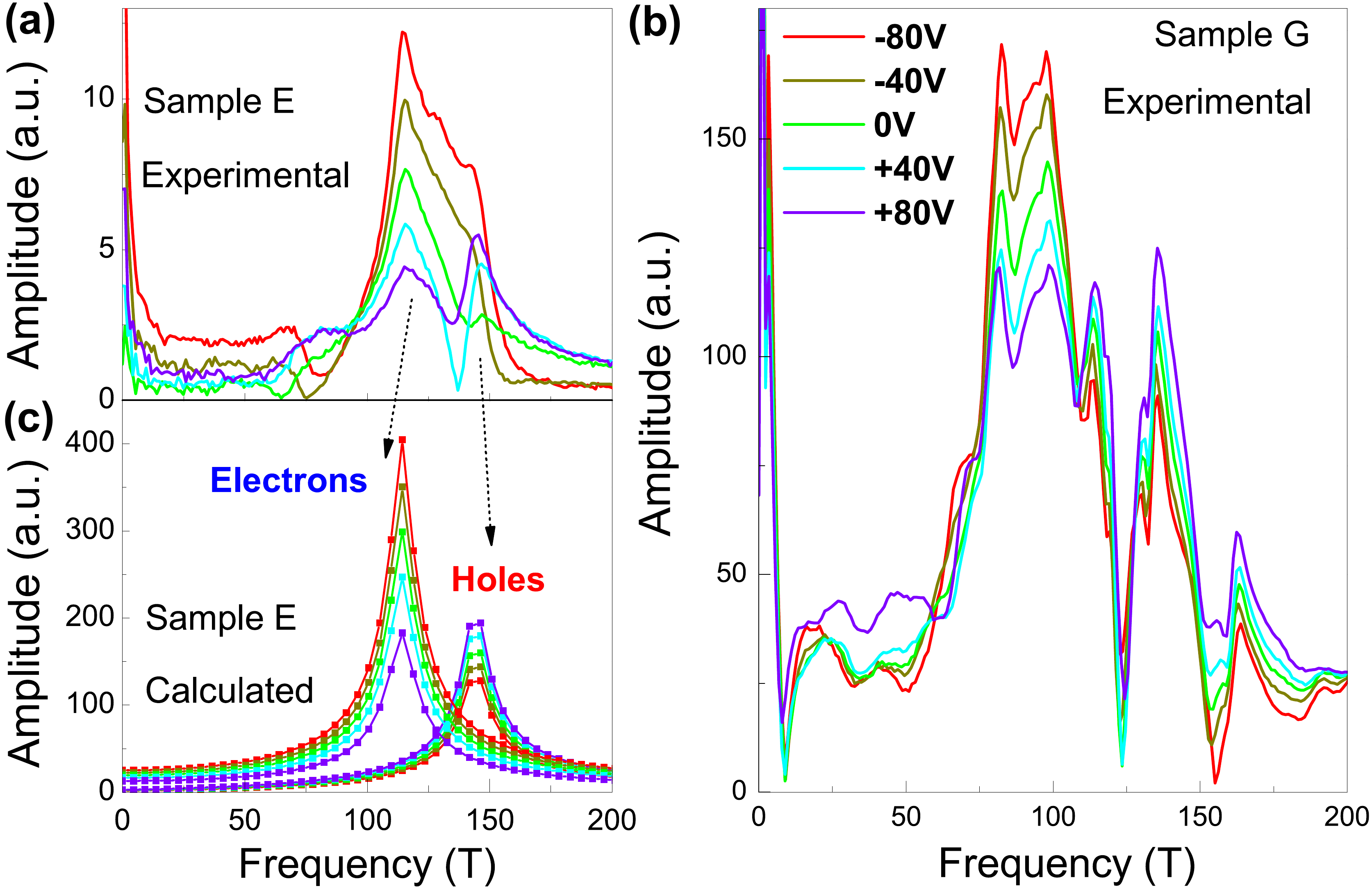}
\caption{\label{figS3} (Color online) \textbf{Gate-dependent Shubnikov-de Haas (SdH) oscillations.} \textbf{(a-b)} Fourier spectrum of the SdH resistivity oscillations $\Delta\rho_{osc}$ measured in Sample E (10 nm thick) and in Sample G (41 nm thick) for different values of gate voltage $V_g$ and at $T=250$ mK. \textbf{(c)} Fourier spectrum of the SdH oscillation $\Delta\rho_{osc}^{calc}$ calculated using Eq. (5) in the main text with the $V_g$-dependent values of electron and hole mobility extracted from the analysis of classical magnet-transport for Sample E (curves of a same color in all panels correspond to the same $V_g$ value, as indicated by the legend in panel (b)).}
\end{figure}

\end{document}